%% file: main.tex
\documentclass[twocolumn,amsmath,amssymb]{revtex4-1}
\usepackage{bm}
\usepackage{braket}
\usepackage[pdftex]{graphicx}

\begin{document}

\title{
Self-consistent Analysis of Doping Effect for Magnetic Ordering in Stacked-Kagome Weyl System
}

\author{Akihiro Ozawa$^1$}\thanks{akihiro.ozawa.s4@dc.tohoku.ac.jp}
\author{Kentaro Nomura$^{1,2}$}\thanks{kentaro.nomura.e7@tohoku.ac.jp}

\affiliation{$^1$Institute for Materials Research, Tohoku University, Katahira, Aoba-ku, Sendai 980-8577}
\affiliation{$^2$Center for Spintronics Research Network, Tohoku University, Katahira, Aoba-ku, Sendai 980-8577
}

\begin{abstract}
We theoretically study the carrier doping effect for magnetism in a stacked-kagome system $\rm{{Co}_3{Sn}_2{S}_2}$ based on an effective model and the Hartree-Fock method.
We show the electron filling and temperature dependences of the magnetic order parameter.
The perpendicular ferromagnetic ordering is suppressed by hole doping, wheres undoped $\rm{{Co}_3{Sn}_2{S}_2}$ shows magnetic Weyl semimetal state.
Additionally, in the electron-doped regime, we find a non-collinear antiferromagnetic ordering. 
Especially, in the non-collinear antiferromagnetic state, by considering a certain spin-orbit coupling, the finite orbital magnetization and the anomalous Hall conductivity are obtained.
\end{abstract}

\maketitle

\section{Introduction}

Magnetic kagome-lattice systems such as $\rm{{Mn}_3{Sn}}$\cite{Nakatsuji2015,Suzuki2017,Balents2017,Ito2017,SSZhang2020}, $\rm{{Fe}_3{Sn}_2}$\cite{Ye2018,Yin2018}, and $\rm{{Co}_3{Sn}_2{S}_2}$\cite{Liu2018,Xu2018,Wang2018,Liu2019,tanaka2020}~(CSS)  are attracting a great deal of attentions because of their diverse electronic and magnetic properties. 
The anomalous Hall effect, originated from the topological gapless points in momentum space called the  Weyl points\cite{Wan2011,Burkov2011,Armitage2018}, is one of the significant transport properties in these materials. 
Especially, CSS possesses the small Fermi surface with the Weyl points and is called the Weyl semimetal\cite{Liu2018}. 
In addition to the electronic properties, these  systems show different magnetic ordering, although they commonly have kagome-lattice layers\cite{Barros2014}.
$\rm{{Mn}_3{Sn}}$ shows a non-collinear antiferromagnetic~(AF) arrangement in which the magnetic moments of Mn are oriented at a relative angle of $120^{\circ}$ in the kagome plane\cite{Nakatsuji2015}.
$\rm{{Fe}_3{Sn}_2}$ shows ferromagnetic~(FM) ordering with the in-plane magnetic anisotropy\cite{Ye2018,Yin2018}.
In CSS, although the ground state shows perpendicular FM  ordering\cite{Liu2018,Ikeda2021,Shiogai2021}, recent experiments predict a non-collinear AF  arrangement at finite temperature\cite{Guguchia2020,Guguchia2021,Zhang2021}.
According to the theory of metallic magnetism\cite{Yoshida},  it has been established  that the Fermi surface structure plays an important role for magnetic ordering.
Therefore, it is expected that the magnetic ordering is altered by tuning the Fermi level. 
However, the theoretical investigations for the magnetic ordering with different Fermi levels in stacked-kagome systems are not well achieved. \par
\begin{figure}[t]
  \includegraphics[width=1.0\hsize]{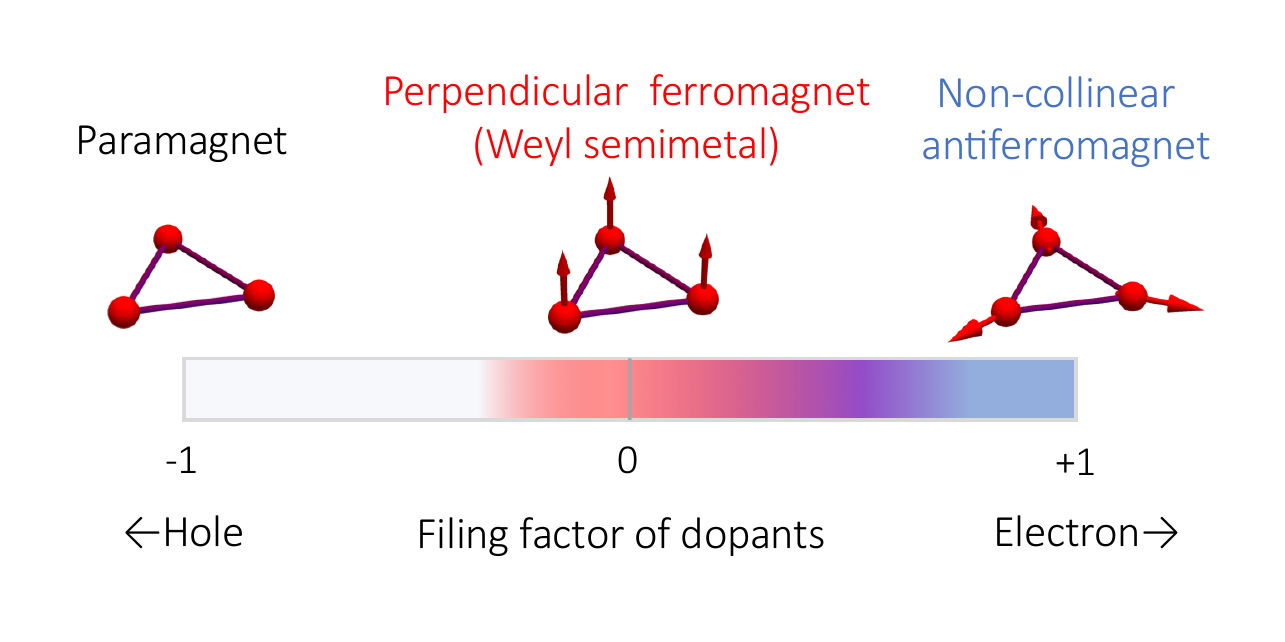}
\caption{
Possible phases in  doped  $\rm{{Co}_3{Sn}_2{S}_2}$. 
In undoped $\rm{{Co}_3{Sn}_2{S}_2}$, Weyl semimetal phase with perpendicular ferromagnetic ordering appears. 
In hole-doped $\rm{{Co}_3{Sn}_2{S}_2}$, the ferromagnetic ordering is suppressed and the system becomes paramagnetic. 
In electron-doped $\rm{{Co}_3{Sn}_2{S}_2}$, a non-collinear antiferromagnetic ordering appears.
  }
\label{fig:overview}
\end{figure}

In this paper, based on  the effective model of the magnetic Weyl semimetal CSS\cite{Ozawa2019}, we study the magnetic ordering with respect to the experimentally controllable parameters, the filling factor of dopants and temperature.
Our results for magnetic ordering are summarized as a schematic picture in Fig.~\ref{fig:overview}.
A non-collinear AF ordering appears by electron doping, wheres undoped system shows the perpendicular ferromagnetic Weyl state.  
As characteristic properties in the non-collinear AF state, the orbital magnetization and the anomalous Hall conductivity become finite by considering a certain spin-orbit coupling.

\section{Tight-binding Hamiltonian and Hartree-Fock mean-field formalism} 
\begin{figure*}[t]
  \includegraphics[width=1\hsize]{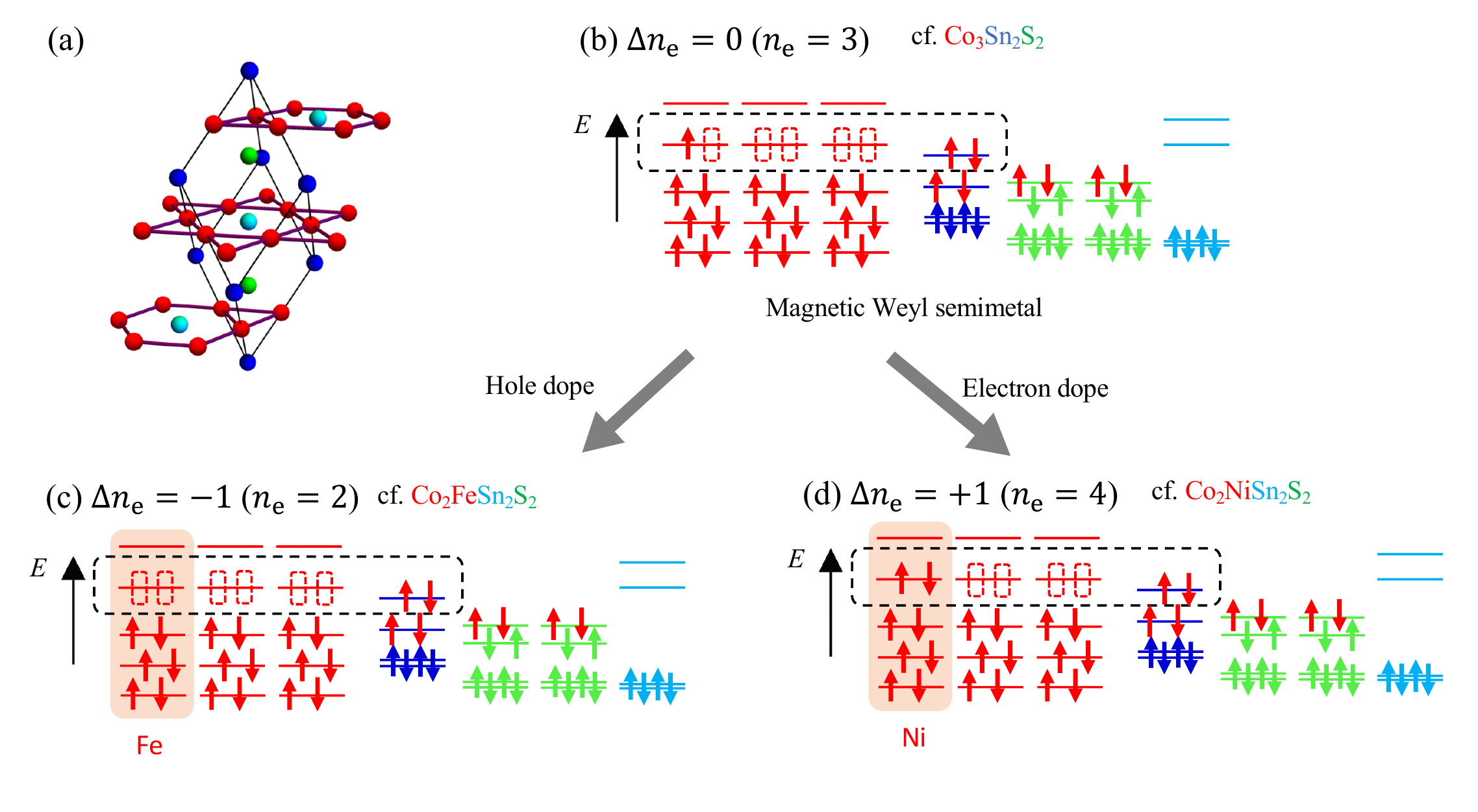}
\caption{
(a)~ Crystal structure of $\rm{{Co}_3{Sn}_2{S}_2}$. 
Co forms the kagome lattice network and sandwiches two layers of triangle lattice formed by Sn and S, respectively.
(b)~Anticipated electronic structures of undoped $\rm{{Co}_3{Sn}_2{S}_2}$.
A dashed box indicates the limited orbitals in our effective model.
The total electron number per unit cell is $n_{\text{e}}=3$. 
(c)~Electronic structure when the total electron number is $n_{\text{e}}=2$. 
In experimental situation, one hole is doped by substituting Co with Fe, in each unit cell.
(d)~Electronic structure when the total electron number  per unit cell is $n_{\text{e}}=4$. 
In experimental situation, one electron is doped by substituting Co with Ni, in each unit cell.
}
\label{fig:doping}
\end{figure*}

First, we briefly introduce the effective model of CSS.
In our previous study\cite{Ozawa2019}, we constructed an effective two-orbital model of CSS, by considering few orbitals.
This model reproduces the electronic band structure which is similar to that obtained by first-principles calculations\cite{Liu2018, Xu2018}.
Figure~\ref{fig:doping}(a) shows the original crystal structure of CSS.
The stackcked kagome layers consist of Co and sandwich two types of triangle layers which consist of Sn and S, respectively.
In the effective model,  one $d$ orbital from Co forming kagome layers and $p$ orbital from interlayer Sn are extracted as a dashed box in Fig.~\ref{fig:doping}(a) shows.
All other orbitals are neglected in the following for simplicity.
The primitive translation vectors are 
$\bm{a}_1=(\frac{a}{2},0,c)$, 
$\bm{a}_2=(-\frac{a}{4},\frac{\sqrt{3}a}{4},c)$, 
$\bm{a}_3=(-\frac{a}{4},-\frac{\sqrt{3}a}{4},c)$.
In the following we  set $c=\frac{\sqrt{3}a}{2}$ for simplicity. 
The hopping term of this model is given by,
\begin{align}
H_{\text{0}}=H_\text{{d-p}}+H_{\rm{KM}}.
\label{eqn:hopping term}
\end{align}
$H_{\text{d-p}}$ is the spin independent hopping term, $H_{\rm{KM}}$ is the spin-orbit coupling term.
First, we explain  $H_\text{{d-p}}$,
\begin{align} \nonumber
 H_\text{{d-p}}=- \sum_{ij\sigma} [t_{ij} d^{\dagger}_{i\sigma} d_{j\sigma}+t^{\rm{dp}}_{ij}(d^{\dagger}_{i\sigma}p_{j\sigma}+p^{\dagger}_{i\sigma}d_{j\sigma})]\\ 
+\epsilon_{\rm{p}}\sum_{i \sigma} p^{\dagger}_{i\sigma}p_{i\sigma}. 
\label{eqn:dphopping}
\end{align} 
$d_{i\sigma}$  and $p_{i\sigma}$ are the annihilation operators of $d$ orbital on the kagome lattice and $p$ orbital on the triangle lattice, respectively.
$t_{ij}$ includes the first and second-nearest neighbor hopping, $t_1$ and $t_2$, in the intra-kagome layer, inter-kagome layer hopping $t_z$.
$t^{\rm dp}$ indicates  $dp$ hybridization between $d$ orbital of Co and $p$ orbital of Sn.
$\epsilon_p$ is the on-site potential of $p$ orbital on Sn.
$H_{\rm{KM}}$ describes the Kane-Mele type SOC term\cite{Kane2005, Guo2009} on the intra kagome layer  given as follows,
\begin{align}
H_{\rm{KM}}=-\text{i}\it{t}_{\rm{KM}} \sum_{\langle \langle ij \rangle \rangle \sigma\sigma'} \nu_{ij} \cdot d^{\dagger}_{i\sigma} {\sigma}^z_{\sigma\sigma'} d_{j\sigma'}.
\label{eqn:kanemele}
\end{align}
$t_{\text{KM}}$ is the hopping strength and the summation $\langle \langle ij \rangle \rangle$ is about intra layer second-nearest-neighbor sites.
The sign is $\nu_{ij}=+1(-1)$, when the electron moves counterclockwise~(clockwise) to get to the second-nearest-neighbor site on the kagome plane\cite{Kane2005, Guo2009}.
Spin-orbit coupling plays a role to obtain the Weyl points\cite{Liu2018,Ozawa2019}.\par 
Next, we construct the mean-field Hamiltonian  by using the Hartree-Fock approximation.
In order to discuss the itinerant magnetism due to the electron correlation, we introduce the on-site Coulomb interaction term.
The on-site Coulomb interaction terms for $d$ orbital $H_{dd}^{\rm{U}}$ and $p$ orbital $H_{pp}^{\rm{U}}$ are respectively given by, 
\begin{eqnarray} 
H_{dd}^{\rm{U}}= U_{dd}\sum_{i}\sum_{\alpha} d^{\dagger}_{i\alpha\uparrow}d^{\dagger}_{i\alpha\downarrow} d_{i\alpha\downarrow}d_{i\alpha\uparrow},
\label{eqn:coulomb_d}
\end{eqnarray}
\begin{eqnarray} 
H_{pp}^{\rm{U}}=U_{pp}\sum_{i} p^{\dagger}_{i\uparrow}p^{\dagger}_{i\downarrow} p_{i\downarrow}p_{i\uparrow}.
\label{eqn:coulomb_p}
\end{eqnarray}
$U_{dd}$ and $U_{pp}$ are the bare on-site Coulomb interaction strengths of $d$ orbital on Co and of $p$ orbital on Sn, respectively.
$i$ and $\alpha=\rm{A, B,~or~ C} $  indicate the position of the unit cell and  the sublattice index of Co, respectively. 
We assume that the fluctuation of the magnetic moment is small.
Thus we introduce the Hartree-Fock approximation $H^{U}_{dd}\sim H^{HF}_{dd}$, $H^{U}_{pp}\sim H^{HF}_{pp}$ for the two-body operators  in Eq.~(\ref{eqn:coulomb_d}) and Eq.~(\ref{eqn:coulomb_p}) as,
\begin{align} \nonumber
&H_{dd}^{\rm{HF}} = U_{dd}\sum_{i\alpha}\bigl[ \langle n_{i\alpha\uparrow}\rangle n_{i\alpha\downarrow}+\langle n_{i\alpha\downarrow}\rangle n_{i\alpha\uparrow}-\langle n_{i\alpha\uparrow}\rangle\langle n_{i\alpha\downarrow}\rangle \\ \nonumber
&~~~~~~~~~~~~-\langle d^{\dagger}_{i\alpha\uparrow}d_{i\alpha\downarrow}\rangle d^{\dagger}_{i\alpha\downarrow}d_{i\alpha\uparrow} -\langle d^{\dagger}_{i\alpha\downarrow}d_{i\alpha\uparrow}\rangle d^{\dagger}_{i\alpha\uparrow}d_{i\alpha\downarrow} \\
&~~~~~~~~~~~~~~~~~~~~~~+\langle d^{\dagger}_{i\alpha\uparrow}d_{i\alpha\downarrow}\rangle\langle d^{\dagger}_{i\alpha\downarrow}d_{i\alpha\uparrow}\rangle \bigr]
\label{eqn:HF_d},
\end{align}
\begin{eqnarray}
H^{\rm{HF}}_{pp}= U_{pp}\sum_{i}\bigl[ \langle n_{ip\uparrow}\rangle n_{ip\downarrow}
+\langle n_{ip\downarrow}\rangle n_{ip\uparrow} 
-\langle n_{ip\uparrow}\rangle\langle n_{ip\downarrow}\rangle  \bigr]. \nonumber \\
\label{eqn:HF_p} 
\end{eqnarray}
$n_{i\alpha\sigma}=d^{\dag}_{i\alpha\sigma}d_{i\alpha\sigma}$ and $n_{ip\sigma}=p^{\dag}_{i\sigma}p_{i\sigma}$ are the particle number operators of Co and Sn, with spin $\sigma$ on $i$th unit cell, respectively.
We neglect the in-plane component of magnetization on Sn site for simplicity.
The total mean-field Hamiltonian $ H_{\rm{MF}}$ is given by,
 \begin{eqnarray}
H_{\rm MF} &=& H_{\rm{0}}+H_{dd}^{\rm{HF}}+H_{pp}^{\rm{HF}}
\label{eqn:MFT}.
\end{eqnarray}
We assume that the translational symmetry of the crystal structure remains even in the magnetically ordered phase.
The mean-field Hamiltonian in momentum space can be obtained by using the Fourier transformation $d_{i\alpha\sigma}=\frac{1}{\sqrt{N}} \sum_{\bm{k}} \mathrm{e}^{i\bm{k} \cdot \bm{R}_i}d_{\bm{k}\alpha\sigma}$,  $p_{i\sigma}=\frac{1}{\sqrt{N}} \sum_{\bm{k}} \mathrm{e}^{i\bm{k} \cdot \bm{R}_i}p_{\bm{k}\sigma}$.
Here $\bm{k}$ is the crystal momentum and $N$ is the number of unit cells.
The Bloch Hamiltonian matrix $ \mathcal{H}_{\text{MF}}(\bm{k})$ can be written in the form,
$H_{\text{MF}}
=\sum_{\bm{k},\sigma} C^{\dagger}_{\bm{k}\sigma} \mathcal{H}_{\text{MF}}(\bm{k}) C_{\bm{k}\sigma},$
where
$C_{\bm{k}\sigma}^{\dag}=(d^{\dag}_{\bm{k}A\sigma}, d^{\dag}_{\bm{k}B\sigma}, d^{\dag}_{\bm{k}C\sigma}, p^\dag_{\bm{k}\sigma})$ and $\mathcal{H}_{\text{MF}}(\bm{k})$ is given by $8 \times 8$ matrix,
\begin{align}
\mathcal{H}_{\text{MF}}(\bm k)=\mathcal{H}_{\text{0}}(\bm k)+\mathcal{H}_{\text{exc}}+H_E
\label{eqn:MFT_k},
\end{align}
in momentum space.
$\mathcal{H}_{\rm{exc}}$ is the exchange term which describes coupling between the mean-field parameter and  spins of electrons as,
\begin{align}
\mathcal{H}_{\rm{exc}}=-\frac{U}{2}{\rm diag}[ {\bm\sigma}\cdot \langle {\bm m}_A \rangle , \bm{\sigma} \cdot\langle \bm{m}_B \rangle ,  \bm{\sigma}\cdot \langle \bm{m}_C \rangle , \sigma_z \langle m^z_S \rangle].
\label{eqn:exc}
\end{align}
${\bm\sigma}$ is the vector of Pauli matrices which corresponds to the spin of electron. 
$\langle {\bm m}_{\alpha} \rangle$ and $\langle m^{z}_S \rangle$  are the mean-field parameters on the $\alpha$ sublattice of Co and Sn, respectively.
In this mean-field Hamiltonian Eq.~(\ref{eqn:MFT_k}), the $z$-component of magnetization and particle number  on each site are computed as  
$\langle m^z_{\gamma}\rangle=\langle n_{\gamma\uparrow}\rangle -\langle n_{\gamma\downarrow}\rangle$,
$\langle n_{\gamma}\rangle=\langle n_{\gamma\uparrow}\rangle +\langle n_{\gamma\downarrow}\rangle$.
Here, we use the simplified sublattice index as $\gamma \in \alpha,S$, and  $\langle n_{\gamma\sigma}\rangle =\frac{1}{N}\sum_{\lambda,\bm{k}}\langle \lambda,\bm{k}|P^{\gamma}_{\sigma}|\lambda,\bm{k}\rangle f(E_{\lambda\bm{k}}-\mu)$.
In-plane components can be obtained as, 
$\langle m^x_{\alpha}\rangle=2 {\rm Re}\langle d_{\alpha \uparrow}^{\dagger} d_{\alpha\downarrow}\rangle$,
$\langle m^y_{\alpha}\rangle=2 {\rm Im}\langle d_{\alpha \uparrow}^{\dagger} d_{\alpha\downarrow}\rangle$, where  
$\langle d_{\alpha \uparrow}^{\dagger} d_{ \alpha\downarrow}\rangle=\frac{1}{N}\sum_{\lambda,\bm{k}}\langle \lambda,\bm{k}|P^{\alpha}\sigma^{+}|\lambda,\bm{k}\rangle f(E_{\lambda\bm{k}}-\mu)$.
$f(E_{\lambda\bm{k}})$ is the Fermi-Dirac distribution function.
$\mu$ is the chemical potential and discussed in detail in the next section.
$P^\gamma$ is the projection operators for $\gamma$ site with spin $\sigma$.
$\sigma^{+}$ is given by   $\sigma^{+}=\sigma_x+i\sigma_y$.
Third term $H_E$ is given by, 
\begin{align} \nonumber
& H_E= \frac{U_{dd}}{4}{\rm diag} [ E_A ,  E_B, E_C, 0]+\frac{U_{pp}}{4}{\rm diag} [ 0 ,  0, 0, E_S] \\ 
&+\frac{U_{dd}}{2} {\rm diag}[\langle n_A \rangle, \langle n_B \rangle ,\langle n_C\rangle,0 ]
+\frac{U_{pp}}{2} {\rm diag}[0, 0 ,0,\langle n_S \rangle ]
\label{eqn:energy}
\end{align} 
$E_\alpha=\langle \bm{m}_\alpha \rangle^2-\langle n_\alpha \rangle^2$ and $E_S=\langle m^z_S \rangle^2-\langle n_S \rangle^2$.
For each $\bm{k}$, the Bloch state is given as an eight component vector $\ket{\lambda, \bm{k}}$, where $\lambda$ is the band index.
$E_{\lambda\bm{k}}$ is the eigenvalue of $\ket{\lambda, \bm{k}}$.
The eigenvector $|\lambda,\bm{k}\rangle$ and order parameters $\langle {\bm m}_{\alpha}\rangle$ can be obtained by diagonalizing  $\mathcal{H}_{\text{MF}}(\bm k)$ so that the  Eq.~(\ref{eqn:MFT_k}) should be calculated  self-consistently.
In the following, we set $t_1$ as a unit of energy, $t_2=0.6t_1$, $t_{\rm{dp}}=2.35t_1$, $t_{\rm z}=-1.2t_1$, $\epsilon_p=-8.5t_1$, $t_{\rm KM}=0.2t_1$, $U_{dd}=7.0t_1$, and  $U_{pp}=5.5t_1$.
These parameters are chosen to fit the band structure to the result obtained by first-principles calculations\cite{Liu2018, Luo2021,Yanagi2021}.

\begin{figure*}[t]
  \includegraphics[width=1\hsize]{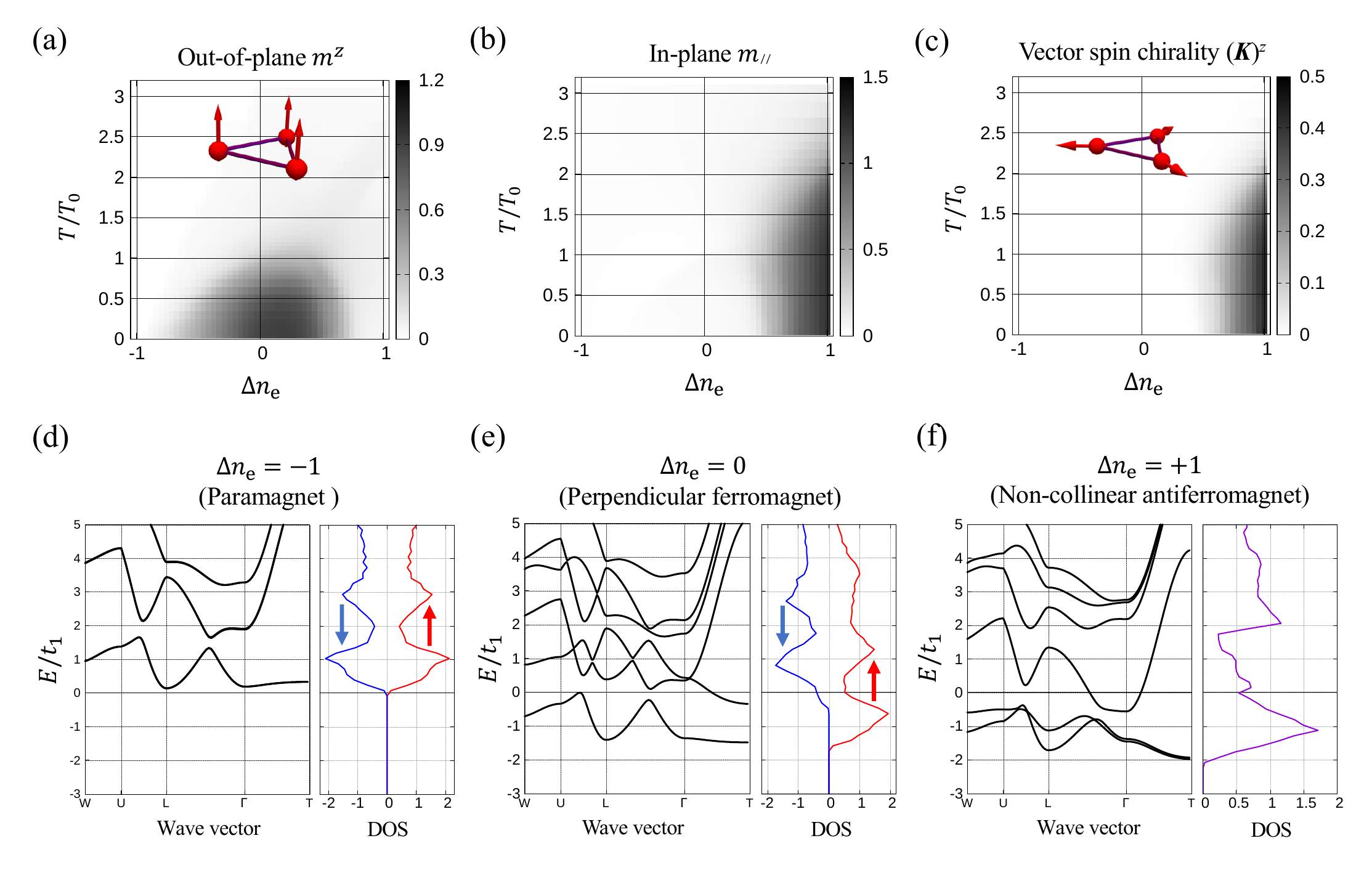}
\caption{
Color maps for 
(a)~$z$ component of magnetization $m^z$ per unit cell in units of $\mu_{\rm B}$, 
(b)~in-plane component of magnetization and  
(c)~$z$ component of vector spin chirality with respect to the filling factor of dopants and temperature.
In case $\Delta n_{\text{e}}\sim0.0$, 
(a)~ferromagnetic ordering with $m^z\sim 0.9$ appears.
$m^z$ decreases as $\Delta n_{\rm e}$ deviates from $\Delta n_{\text{e}}\sim0$.
When $\Delta n_{\text{e}}\sim+1.0$, 
(a)~$z$ component of magnetization diminishes, while 
(b)~in-plane component of magnetization and 
(c)~$z$ component of the vector spin chirality become finite, indicating non-collinear antiferromagnetic state.
Electronic band structures of (d)~paramagnetic state, (e)~perpendicular ferromagnetic state and (f)~non-collinear antiferromagnetic state obtained by the Hartree-Fock method.
In (d)~$\Delta n_{\rm e}=-1$, system is paramagnetic and the chemical potential is close to the band gap.
In (e)~$\Delta n_{\rm e}=-1$, system is ferromagnetic, the chemical potential is located near the local minimum of the spin majority band, corresponding to the Weyl points, and near the gap of the spin minority band.
In (f) non-collinear antiferromagnetic state, the electronic band dispersion around the L point remains almost unchanged, comparing to that in ferromagnetic state.
}
\label{fig:order}
\end{figure*}

\section{Condition of total number of electrons in unit cell}

Next, we discuss the chemical potential in our theoretical model.
In the following, we assume that the doping effect is considered as only a change of the number of electron per unit cell, and the randomness due to the impurities is neglected.
As mentioned in the previous section, we extracted one orbital from five $d$ orbitals of each Co and one orbital from $p$ orbitals of interlayer Sn, and neglected all other orbitals as shown in Fig.~\ref{fig:doping}(a).
Therefore, the unit cell has (3+1)$\times$ 2 =8 states including the spin degrees of freedom in our model.
To determine  $\mu$ appropriately, we discuss the electronic orbital configurations in the doped CSS.
As discussed in our previous paper\cite{Ozawa2019}, in the undoped CSS, we assume that one of three sites of Co is occupied by one electron, and interlayer Sn site is occupied by two electrons. 
Thus the total number of electrons in limited orbitals, is $n_\text{e}=3$  per unit cell as shown in Fig.~\ref{fig:doping}(a).
This configuration is consistent with the magnetization per unit cell $m_z\sim 1.0$  as obtained by  experiment\cite{Liu2018}. 
In this work, we study the doping effect to the undoped CSS.
To clearly characterize the filling factor of dopants, we use $\Delta n_\text{e}$ as the deviation from $n_\text{e}=3$ in the following results.
Therefore, $n_\text{e}=3$ is equivalent to $\Delta n_\text{e}=0$.
When one Co in each unit cell is substituted with one Fe, the anticipated electronic orbital configuration is shown in Fig.~\ref{fig:doping}(c).
In this case, the total number of electrons per unit cell is $n_\text{e}=2$ so $\Delta n_\text{e}=-1$.
Presumably, even if Sn is substituted with In, instead of substituting Co with Fe, the total number of electrons per unit cell is same as that in Fig.~\ref{fig:doping}(c).
This is because one electron at the Co orbital is expected to move to the In orbital, which is assumed to be energetically low. 
On the other hand, when one of Co site in each unit cell is substituted with one Ni, the anticipated electronic orbital configuration is shown in Fig.~\ref{fig:doping}(d).
In this case, the total number of electrons per unit cell is $n_\text{e}=4$ so $\Delta n_\text{e}=+1$.
The chemical potential $\mu$ is numerically determined to satisfy the following equation,
\begin{align}
n_\text{e}=\int^{\infty}_{-\infty}d\epsilon \rho(\epsilon)f(\epsilon-\mu, T).
\label{eqn:chem_pot}
\end{align}
Here, $\rho(\epsilon)$ is the density of states per unit cell, $k_{\rm{B}}$ is the Boltzman constant and $T$ is temperature.
According to the above argument, we can determine the chemical potential $\mu$ using the Eq.~(\ref{eqn:chem_pot}).

\section{Magnetic ordering}

Next, we investigate the magnetic ordering with respect to the filling factor of dopants $\Delta n_\text{e}$ and temperature $T$.
In Fig.~\ref{fig:order}, the $\Delta n_\text{e}\text{-}T$ dependence of 
(a)~the $z$ component of magnetization  ${m^z}=\sum_{\gamma}\langle m^{z}_{\gamma}\rangle$, 
(b)~the in-plane component of magnetization $m_{//\!}=\sum_{\alpha}\sqrt{\langle m^{x}_{\alpha}\rangle^2+\langle m^y_{\alpha} \rangle^2}$~($\alpha$=A, B, and C), 
(c) the $z$-component of the vector spin chirality\cite{Grohol2005}  $({\bm K})^z=({\bm S}_A \times {\bm S}_B+{\bm S}_B \times {\bm S}_C+{\bm S}_C \times {\bm S}_A)^z$ are shown.
Additionally, in Fig.~\ref{fig:order}, the band structure and the density of states at 
(d)~$\Delta n_\text{e}=-1$, 
(e)~$\Delta n_\text{e}=0$, and 
(f)~$\Delta n_\text{e}=+1$  are shown.
First, we study the FM ordering with the perpendicular anisotropy in undoped CSS~($\Delta n_{\rm e}=0$).
Figures~\ref{fig:order}~(a) and \ref{fig:order}~(b) show, at low temperature, $m^z \sim 0.9$ and $m_{//\!}\sim 0$ in undoped case~($\Delta n_\text{e}=0$), indicating FM ordering with the  perpendicular anisotropy.
The value $m^z\sim 0.9$ is consistent with results obtained by first-principles calculations\cite{Liu2018} and experiment\cite{Kassem2017,Liu2018}.
We find the critical temperature in undoped case  being $T_0= 0.4t_1/k_{\rm B}$.
The band structure and the density of states in undoped case~($\Delta n_\text{e}=0$) obtained by the Hartree-Fock method are shown in Fig.~\ref{fig:order}(e).
We set  $k_{\text{B}}T/t_1=0.01$.
$E_{1}/t_1=0$ is set as the chemical potential $\mu$ obtained by Eq.~(\ref{eqn:chem_pot}).
We do not depict the lower two bands because they are energetically apart from $\mu$. 
As  the right panel of Fig.~\ref{fig:order}(e) shows, near $\mu$, the spin up band has a relatively small density of states corresponding to the Weyl points.
Whereas the spin down band is close to  the band gap.
This describes the spin-polarized Weyl semimetalic state in undoped CSS. \par
Next, we show the suppression of the FM ordering in the hole-doped regime.
Figure~\ref{fig:order}(a) shows that the FM transition temperature  decreases when $\Delta n_\text{e}<0$.
This suppression of FM ordering by hole-doping is consistent with experiment in ${\rm Co}_{3-x}{\rm Fe}_x{\rm Sn}_2{\rm S}_2$\cite{Weihrich2006,Kassem2016,Huibin2020,Shen2020} and first-principles calculations and experiment for  ${\rm Co}_{3}{\rm In}_{x}{\rm Sn}_{2-x}{\rm S}_2$\cite{Kassem2015,Yanagi2021}.
The non-magnetic band structure and the density of states in the hole-doped CSS when $\Delta n_{\rm e}=-1$ are shown in Fig. 3(d).
In this situation, $\mu$ is close to the band gap, indicating a paramagnetic state with small carriers. \par
Then, we study the electron-doped regime.
This situation could be realized experimentally in ${\rm Co}_{3-x}{\rm Ni}_{x}{\rm Sn}_2{\rm S}_2$ \cite{Kubodera2006, Thakur2020}. 
As shown in Fig.~\ref{fig:order}(a), $m^z$ decreases as $\Delta n_{\rm e}$ increases from $\Delta n_{\rm e}=0$.
As Fig.~\ref{fig:order}(c) shows, $z$ component of vector spin chirality  becomes positive as $\Delta n_{\rm e}$ increases, while  $m_z$ vanishes.
Especially, when $\Delta n_\text{e}=+1$, we find that the spin configuration becomes as  ${\bm m}_{A}=m(1,0,0)$,  ${\bm m}_{B}=m(\cos{(2\pi/3)},\sin{(2\pi/3)},0)$, ${\bm m}_{C}=m(\cos{(4\pi/3)},\sin{(4\pi/3)},0)$, where $ m\sim 0.5 \mu_{\rm B}$.
These results conclude that the non-collinear AF ordering appears within the restricted order parameter space of our model.
In Fig.~\ref{fig:order}(f), the electronic band structure and the density of states in the non-collinear AF state are shown.
Around the L point, the  band dispersion  near $\mu$ remains almost unchanged from that in FM state~[Fig.~\ref{fig:order}(e)].
In Fig.~\ref{fig:order}(c), the non-collinear AF ordering sustains up to $T/T_0 \sim 2.3$.
However, we note that the magnetic transition temperature is overestimated due to the use of Hartree-Fock method\cite{Dai2005}.
On the other hand, at low temperature the appearance of magnetic ordering is reliable.

\section{orbital magnetization in antiferromagnetic state}
In the previous section, we showed that the non-collinear AF ordering appears in the electron-doped regime. 
Here, we discuss the orbital magnetization and the anomalous Hall conductivity, characterizing the non-collinear AF state.
Considering a certain additional SOC, the orbital magnetization and the anomalous Hall conductivity become finite in the non-collinear AF state.
We note that, by considering only the intralayer Kane-Mele SOC given by Eq.~(\ref{eqn:kanemele}), both of these values vanish.
As an additional interaction, we introduce the interlayer Kane-Mele type SOC due to the honeycomb structure.
\begin{align}
 H^{z}_{\rm KM }=-{\rm i}\it{t}^{z}_{\rm KM } \sum_{\langle \langle ij \rangle \rangle \sigma\sigma'} {\bm \eta}_{ij} \cdot d^{\dagger}_{i\sigma} {\bm\sigma}_{\sigma\sigma'} d_{j\sigma'}.
 \label{eqn:inter_kanemele}
\end{align}
Here, ${\bm \eta}_{ij}$ are given by ${\bm \eta}_{CA}=\frac{\bm{a}_1}{2} \times \frac{\bm{a}_3}{2}$,  ${\bm \eta}_{AB}=\frac{\bm{a}_2}{2} \times \frac{\bm{a}_1}{2}$, and  ${\bm \eta}_{BC}=\frac{\bm{a}_3}{2} \times \frac{\bm{a}_2}{2}$.
Although the magnetic ordering remains mostly unchanged by this additional SOC Eq.~(\ref{eqn:inter_kanemele}), this term makes the orbital magnetization and the AHC finite in non-collinear AF state. \par
We study the spin-moment angle dependences of the orbital magnetization.
The orbital magnetization can be obtained by the formula\cite{Xiao2010, Ceresoli2006, Ito2017,Ominato2019}, 
\begin{figure}[t]
  \includegraphics[width=1.0\hsize]{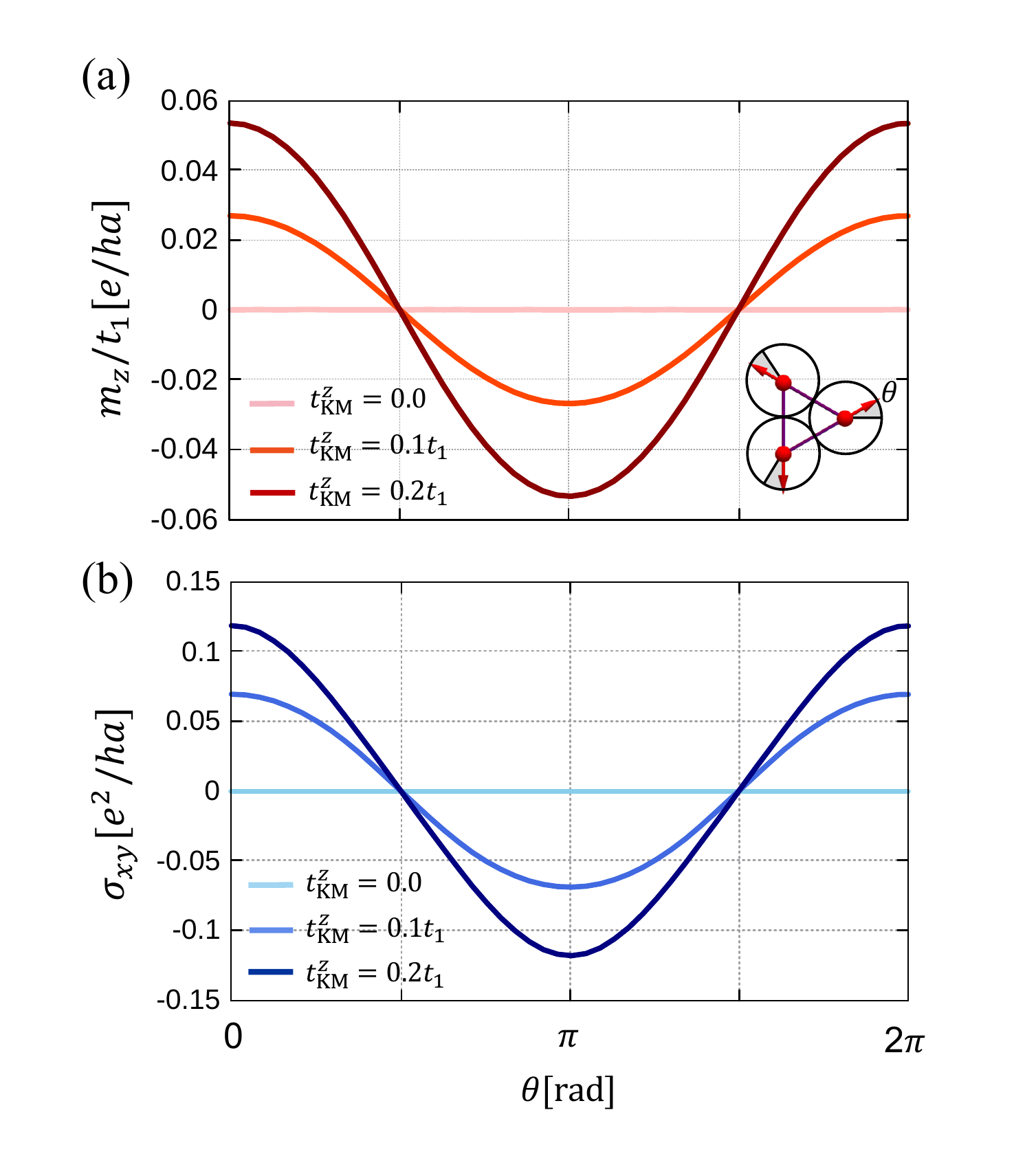}
\caption{
(a)~Orbital magnetizations and (b)~anomalous Hall conductivity for $t^z_{\rm KM}$=$0.0$, $0.1t_1$, and $0.2t_1$, as a function of magnetization angle $\theta$ depicted in an inset of (a).
}
\label{fig:orb_mag}
\end{figure}
\begin{align} \nonumber
&M^{\rm orb}_{\alpha}=\frac{e}{2\hbar}\sum_{\lambda} \int_{\rm BZ} \frac{d^3k}{(2\pi)^3}f_{\lambda{\bm k}}\epsilon_{\alpha\beta\gamma} \\ \nonumber
&~~~~~~~~~~~\times {\rm Im}\sum_{\lambda'\neq\lambda}\frac{\langle \lambda, \bm{k}|\hbar v_{\beta}|\lambda',\bm{k}\rangle \langle \lambda', \bm{k}|\hbar v_{\gamma}|\lambda,\bm{k}\rangle}{(E_{\lambda' {\bm k}}-E_{\lambda {\bm k}})^2}  \\
&~~~~~~~~~~~~~~~~~~\times (E_{\lambda' {\bm k}}+E_{\lambda {\bm k}}-2E_{\rm F}).
\label{eqn:kubo_om}
\end{align}
Here, $v_i~(i = x, y, z)$ is the velocity operator given by $v_{i}= \frac{1}{\hbar} \frac{ \partial H({\bm k})}{\partial k_i}$.
The eigenstates $|\lambda,\bm{k}\rangle$ are obtained by diagonalizing $\mathcal{H}_{\text{0}}(\bm k)+\mathcal{H}_{\text{exc}}$ with Eq.~(\ref{eqn:inter_kanemele}).
Figure \ref{fig:orb_mag}(a) shows $M^{\rm orb}_z$ as a function of the angle of magnetic moment on kagome lattice for $t^{z}_{\rm KM}=0.0$, $0.1t_1$ and $0.2t_1$,. 
Each magnetic moment is rotated with an equivalent relative angle as shown in an inset of Fig.~\ref{fig:orb_mag}(a). 
$E_{\rm F}$ in Eq.~(\ref{eqn:kubo_om}) is obtained by $\Delta n_{\rm e} = +1$ condition and the magnetic order parameters on each site are obtained by Hartree-Fock method.
$M^{\rm orb}_z$ is finite and changes like a $\cos{\theta}$ function.
We note that $M^{\rm orb}_x = M^{\rm orb}_y = 0$. 
These results indicate that our model in the non-collinear AF state shows a finite orbital magnetization although the net magnetization vanishes. 
The direction of the spin moments can be changed by an external magnetic field as similarly discussed in Ref.~\cite{Ito2017}. 
In the presence of an external magnetic field pointing the $z$ direction $B_z$, the orbital magnetization $M^{\rm orb}_z$ couples as $-M^{\rm orb}_zB_z$. 
When the external magnetic field points $+z$ direction, the spin angle $\theta=0$ is energetically favored. 
On the other hand, when the external magnetic field points $-z$ direction, the spin angle $\theta=\pi$  is energetically favored. \par
The change of the spin direction is related to the AHE. 
The intrinsic AHC $\sigma_{xy}$  can be calculated by the formula\cite{Nagaosa2010} given by,
\begin{align} \nonumber
&\sigma_{xy}=e^2\hbar\sum_{\lambda} \int_{\rm BZ} \frac{d^3k}{(2\pi)^3}f_{\lambda{\bm k}} \\
&~~~~~~~~\times {\rm Im}\sum_{\lambda'\neq\lambda}\frac{\langle \lambda, \bm{k}| v_x|\lambda',\bm{k}\rangle \langle \lambda', \bm{k}|v_y|\lambda,\bm{k}\rangle}{(E_{\lambda' {\bm k}}-E_{\lambda {\bm k}})^2}  
\label{eqn:kubo_ahc}.
\end{align}
As shown in Fig.~\ref{fig:orb_mag}(b),  the angle dependence of the AHC is similar to that of the orbital magnetization in Fig.~\ref{fig:orb_mag}(a). 
Therefore, the sign of the AHC changes when the direction of spin moments is changed  by an external magnetic field.
Although the AHC in $\Delta n_{\rm e}=+1$ is smaller than that in ferromagnetic Weyl state~($\Delta n_{\rm e} = 0$) \cite{Ozawa2019}, the change of the direction of spin moments in non-collinear AF state  might be detected by applying a uniform magnetic field. \par
\section{Conclusion}
In this paper, we investigated the magnetic ordering in an effective model of stacked-kagome lattice system CSS, based on the Hartree-Fock method.
We showed the suppression of the perpendicular ferromagnetic ordering by hole doping.
Non-collinear AF phase appears in electron-doped regimes and possesses finite orbital magnetization and the AHC by considering the interlayer SOC.\par
\section{Acknowledgement}
We thank Y. Araki, K. Kobayashi, Y. Motome, A. Tsukazaki, J. Watanabe for valuable discussions. 
This work was supported by 
JSPS KAKENHI Grants No. 20H01830,
JST CREST Grant No. JPMJCR18T2, 
and GP-Spin at Tohoku University.

\input{main.bbl}

\end{document}

%% file: main.bbl
%

%% file: main.bbl
\begin{thebibliography}{41}%
\makeatletter
\providecommand \@ifxundefined [1]{%
 \@ifx{#1\undefined}
}%
\providecommand \@ifnum [1]{%
 \ifnum #1\expandafter \@firstoftwo
 \else \expandafter \@secondoftwo
 \fi
}%
\providecommand \@ifx [1]{%
 \ifx #1\expandafter \@firstoftwo
 \else \expandafter \@secondoftwo
 \fi
}%
\providecommand \natexlab [1]{#1}%
\providecommand \enquote  [1]{``#1''}%
\providecommand \bibnamefont  [1]{#1}%
\providecommand \bibfnamefont [1]{#1}%
\providecommand \citenamefont [1]{#1}%
\providecommand \href@noop [0]{\@secondoftwo}%
\providecommand \href [0]{\begingroup \@sanitize@url \@href}%
\providecommand \@href[1]{\@@startlink{#1}\@@href}%
\providecommand \@@href[1]{\endgroup#1\@@endlink}%
\providecommand \@sanitize@url [0]{\catcode `\\12\catcode `\$12\catcode
  `\&12\catcode `\#12\catcode `\^12\catcode `\_12\catcode `\%12\relax}%
\providecommand \@@startlink[1]{}%
\providecommand \@@endlink[0]{}%
\providecommand \url  [0]{\begingroup\@sanitize@url \@url }%
\providecommand \@url [1]{\endgroup\@href {#1}{\urlprefix }}%
\providecommand \urlprefix  [0]{URL }%
\providecommand \Eprint [0]{\href }%
\providecommand \doibase [0]{http://dx.doi.org/}%
\providecommand \selectlanguage [0]{\@gobble}%
\providecommand \bibinfo  [0]{\@secondoftwo}%
\providecommand \bibfield  [0]{\@secondoftwo}%
\providecommand \translation [1]{[#1]}%
\providecommand \BibitemOpen [0]{}%
\providecommand \bibitemStop [0]{}%
\providecommand \bibitemNoStop [0]{.\EOS\space}%
\providecommand \EOS [0]{\spacefactor3000\relax}%
\providecommand \BibitemShut  [1]{\csname bibitem#1\endcsname}%
\let\auto@bib@innerbib\@empty
\bibitem [{\citenamefont {Nakatsuji}\ \emph {et~al.}(2015)\citenamefont
  {Nakatsuji}, \citenamefont {Kiyohara},\ and\ \citenamefont
  {Higo}}]{Nakatsuji2015}%
  \BibitemOpen
  \bibfield  {author} {\bibinfo {author} {\bibfnamefont {S.}~\bibnamefont
  {Nakatsuji}}, \bibinfo {author} {\bibfnamefont {N.}~\bibnamefont {Kiyohara}},
  \ and\ \bibinfo {author} {\bibfnamefont {T.}~\bibnamefont {Higo}},\
  }\href@noop {} {\bibfield  {journal} {\bibinfo  {journal} {Nature}\ }\textbf
  {\bibinfo {volume} {527}},\ \bibinfo {pages} {212} (\bibinfo {year}
  {2015})}\BibitemShut {NoStop}%
\bibitem [{\citenamefont {Suzuki}\ \emph {et~al.}(2017)\citenamefont {Suzuki},
  \citenamefont {Koretsune}, \citenamefont {Ochi},\ and\ \citenamefont
  {Arita}}]{Suzuki2017}%
  \BibitemOpen
  \bibfield  {author} {\bibinfo {author} {\bibfnamefont {M.-T.}\ \bibnamefont
  {Suzuki}}, \bibinfo {author} {\bibfnamefont {T.}~\bibnamefont {Koretsune}},
  \bibinfo {author} {\bibfnamefont {M.}~\bibnamefont {Ochi}}, \ and\ \bibinfo
  {author} {\bibfnamefont {R.}~\bibnamefont {Arita}},\ }\href@noop {}
  {\bibfield  {journal} {\bibinfo  {journal} {Phys. Rev. B}\ }\textbf {\bibinfo
  {volume} {95}},\ \bibinfo {pages} {094406} (\bibinfo {year}
  {2017})}\BibitemShut {NoStop}%
\bibitem [{\citenamefont {Liu}\ and\ \citenamefont
  {Balents}(2017)}]{Balents2017}%
  \BibitemOpen
  \bibfield  {author} {\bibinfo {author} {\bibfnamefont {J.}~\bibnamefont
  {Liu}}\ and\ \bibinfo {author} {\bibfnamefont {L.}~\bibnamefont {Balents}},\
  }\href@noop {} {\bibfield  {journal} {\bibinfo  {journal} {Phys. Rev. Lett.}\
  }\textbf {\bibinfo {volume} {119}},\ \bibinfo {pages} {087202} (\bibinfo
  {year} {2017})}\BibitemShut {NoStop}%
\bibitem [{\citenamefont {Ito}\ and\ \citenamefont {Nomura}(2017)}]{Ito2017}%
  \BibitemOpen
  \bibfield  {author} {\bibinfo {author} {\bibfnamefont {N.}~\bibnamefont
  {Ito}}\ and\ \bibinfo {author} {\bibfnamefont {K.}~\bibnamefont {Nomura}},\
  }\href@noop {} {\bibfield  {journal} {\bibinfo  {journal} {J. Phys. Soc.
  Jpn}\ }\textbf {\bibinfo {volume} {86}},\ \bibinfo {pages} {063703} (\bibinfo
  {year} {2017})}\BibitemShut {NoStop}%
\bibitem [{\citenamefont {Zhang}\ \emph {et~al.}(2020)\citenamefont {Zhang},
  \citenamefont {Ishizuka}, \citenamefont {Zhang}, \citenamefont {Hal\'asz},\
  and\ \citenamefont {Batista}}]{SSZhang2020}%
  \BibitemOpen
  \bibfield  {author} {\bibinfo {author} {\bibfnamefont {S.-S.}\ \bibnamefont
  {Zhang}}, \bibinfo {author} {\bibfnamefont {H.}~\bibnamefont {Ishizuka}},
  \bibinfo {author} {\bibfnamefont {H.}~\bibnamefont {Zhang}}, \bibinfo
  {author} {\bibfnamefont {G.~B.}\ \bibnamefont {Hal\'asz}}, \ and\ \bibinfo
  {author} {\bibfnamefont {C.~D.}\ \bibnamefont {Batista}},\ }\href@noop {}
  {\bibfield  {journal} {\bibinfo  {journal} {Phys. Rev. B}\ }\textbf {\bibinfo
  {volume} {101}},\ \bibinfo {pages} {024420} (\bibinfo {year}
  {2020})}\BibitemShut {NoStop}%
\bibitem [{\citenamefont {Ye}\ \emph {et~al.}(2018)\citenamefont {Ye},
  \citenamefont {Kang}, \citenamefont {Liu}, \citenamefont {von Cube},
  \citenamefont {Wicker}, \citenamefont {Suzuki}, \citenamefont {Jozwiak},
  \citenamefont {Bostwick}, \citenamefont {Rotenberg}, \citenamefont {Bell},
  \citenamefont {Fu}, \citenamefont {Comin},\ and\ \citenamefont
  {Checkelsky}}]{Ye2018}%
  \BibitemOpen
  \bibfield  {author} {\bibinfo {author} {\bibfnamefont {L.}~\bibnamefont
  {Ye}}, \bibinfo {author} {\bibfnamefont {M.}~\bibnamefont {Kang}}, \bibinfo
  {author} {\bibfnamefont {J.}~\bibnamefont {Liu}}, \bibinfo {author}
  {\bibfnamefont {F.}~\bibnamefont {von Cube}}, \bibinfo {author}
  {\bibfnamefont {C.}~\bibnamefont {Wicker}}, \bibinfo {author} {\bibfnamefont
  {T.}~\bibnamefont {Suzuki}}, \bibinfo {author} {\bibfnamefont
  {C.}~\bibnamefont {Jozwiak}}, \bibinfo {author} {\bibfnamefont
  {A.}~\bibnamefont {Bostwick}}, \bibinfo {author} {\bibfnamefont
  {R.}~\bibnamefont {Rotenberg}}, \bibinfo {author} {\bibfnamefont
  {D.}~\bibnamefont {Bell}}, \bibinfo {author} {\bibfnamefont {L.}~\bibnamefont
  {Fu}}, \bibinfo {author} {\bibfnamefont {R.}~\bibnamefont {Comin}}, \ and\
  \bibinfo {author} {\bibfnamefont {J.~G.}\ \bibnamefont {Checkelsky}},\
  }\href@noop {} {\bibfield  {journal} {\bibinfo  {journal} {Nature}\ }\textbf
  {\bibinfo {volume} {555}},\ \bibinfo {pages} {638} (\bibinfo {year}
  {2018})}\BibitemShut {NoStop}%
\bibitem [{\citenamefont {Yin}\ \emph {et~al.}(2018)\citenamefont {Yin},
  \citenamefont {Zhang}, \citenamefont {Li}, \citenamefont {Jiang},
  \citenamefont {Chang}, \citenamefont {Zhang}, \citenamefont {Lian},
  \citenamefont {Xiang}, \citenamefont {Belopolski}, \citenamefont {Zheng},
  \citenamefont {Cochran}, \citenamefont {Xu}, \citenamefont {Bian},
  \citenamefont {Liu}, \citenamefont {Chang}, \citenamefont {Lin},
  \citenamefont {Lu}, \citenamefont {Wang}, \citenamefont {Jia}, \citenamefont
  {Wang},\ and\ \citenamefont {Hasan}}]{Yin2018}%
  \BibitemOpen
  \bibfield  {author} {\bibinfo {author} {\bibfnamefont {J.-X.}\ \bibnamefont
  {Yin}}, \bibinfo {author} {\bibfnamefont {S.~S.}\ \bibnamefont {Zhang}},
  \bibinfo {author} {\bibfnamefont {H.}~\bibnamefont {Li}}, \bibinfo {author}
  {\bibfnamefont {K.}~\bibnamefont {Jiang}}, \bibinfo {author} {\bibfnamefont
  {G.}~\bibnamefont {Chang}}, \bibinfo {author} {\bibfnamefont
  {B.}~\bibnamefont {Zhang}}, \bibinfo {author} {\bibfnamefont
  {B.}~\bibnamefont {Lian}}, \bibinfo {author} {\bibfnamefont {C.}~\bibnamefont
  {Xiang}}, \bibinfo {author} {\bibfnamefont {I.}~\bibnamefont {Belopolski}},
  \bibinfo {author} {\bibfnamefont {H.}~\bibnamefont {Zheng}}, \bibinfo
  {author} {\bibfnamefont {T.}~\bibnamefont {Cochran}}, \bibinfo {author}
  {\bibfnamefont {A.-Y.}\ \bibnamefont {Xu}}, \bibinfo {author} {\bibfnamefont
  {G.}~\bibnamefont {Bian}}, \bibinfo {author} {\bibfnamefont {K.}~\bibnamefont
  {Liu}}, \bibinfo {author} {\bibfnamefont {T.-R.}\ \bibnamefont {Chang}},
  \bibinfo {author} {\bibfnamefont {H.}~\bibnamefont {Lin}}, \bibinfo {author}
  {\bibfnamefont {Z.-Y.}\ \bibnamefont {Lu}}, \bibinfo {author} {\bibfnamefont
  {Z.}~\bibnamefont {Wang}}, \bibinfo {author} {\bibfnamefont {S.}~\bibnamefont
  {Jia}}, \bibinfo {author} {\bibfnamefont {W.}~\bibnamefont {Wang}}, \ and\
  \bibinfo {author} {\bibfnamefont {M.}~\bibnamefont {Hasan}},\ }\href@noop {}
  {\bibfield  {journal} {\bibinfo  {journal} {Nature}\ }\textbf {\bibinfo
  {volume} {562}},\ \bibinfo {pages} {91} (\bibinfo {year} {2018})}\BibitemShut
  {NoStop}%
\bibitem [{\citenamefont {Liu}\ \emph {et~al.}(2018)\citenamefont {Liu},
  \citenamefont {Sun}, \citenamefont {Kumar}, \citenamefont {Muechler},
  \citenamefont {Sun}, \citenamefont {Jiao}, \citenamefont {Yang},
  \citenamefont {Liu}, \citenamefont {Liang}, \citenamefont {Xu}, \citenamefont
  {Sun}, \citenamefont {Kumar}, \citenamefont {Muechler}, \citenamefont {Sun},
  \citenamefont {Jiao}, \citenamefont {Yang}, \citenamefont {Liu},
  \citenamefont {Liang}, \citenamefont {Xu}, \citenamefont {Kroder},
  \citenamefont {S\"u\ss{}}, \citenamefont {Borrmann}, \citenamefont {Shekhar},
  \citenamefont {Wang}, \citenamefont {Wang}, \citenamefont {Wirth},
  \citenamefont {Chen}, \citenamefont {Goennenwein},\ and\ \citenamefont
  {Felser}}]{Liu2018}%
  \BibitemOpen
  \bibfield  {author} {\bibinfo {author} {\bibfnamefont {E.}~\bibnamefont
  {Liu}}, \bibinfo {author} {\bibfnamefont {Y.}~\bibnamefont {Sun}}, \bibinfo
  {author} {\bibfnamefont {N.}~\bibnamefont {Kumar}}, \bibinfo {author}
  {\bibfnamefont {L.}~\bibnamefont {Muechler}}, \bibinfo {author}
  {\bibfnamefont {A.}~\bibnamefont {Sun}}, \bibinfo {author} {\bibfnamefont
  {L.}~\bibnamefont {Jiao}}, \bibinfo {author} {\bibfnamefont {S.-Y.}\
  \bibnamefont {Yang}}, \bibinfo {author} {\bibfnamefont {D.}~\bibnamefont
  {Liu}}, \bibinfo {author} {\bibfnamefont {A.}~\bibnamefont {Liang}}, \bibinfo
  {author} {\bibfnamefont {Q.}~\bibnamefont {Xu}}, \bibinfo {author}
  {\bibfnamefont {Y.}~\bibnamefont {Sun}}, \bibinfo {author} {\bibfnamefont
  {N.}~\bibnamefont {Kumar}}, \bibinfo {author} {\bibfnamefont
  {L.}~\bibnamefont {Muechler}}, \bibinfo {author} {\bibfnamefont
  {A.}~\bibnamefont {Sun}}, \bibinfo {author} {\bibfnamefont {L.}~\bibnamefont
  {Jiao}}, \bibinfo {author} {\bibfnamefont {S.-Y.}\ \bibnamefont {Yang}},
  \bibinfo {author} {\bibfnamefont {D.}~\bibnamefont {Liu}}, \bibinfo {author}
  {\bibfnamefont {A.}~\bibnamefont {Liang}}, \bibinfo {author} {\bibfnamefont
  {Q.}~\bibnamefont {Xu}}, \bibinfo {author} {\bibfnamefont {J.}~\bibnamefont
  {Kroder}}, \bibinfo {author} {\bibfnamefont {V.}~\bibnamefont {S\"u\ss{}}},
  \bibinfo {author} {\bibfnamefont {H.}~\bibnamefont {Borrmann}}, \bibinfo
  {author} {\bibfnamefont {C.}~\bibnamefont {Shekhar}}, \bibinfo {author}
  {\bibfnamefont {C.}~\bibnamefont {Wang}, \bibfnamefont {Z.~Xi}}, \bibinfo
  {author} {\bibfnamefont {W.}~\bibnamefont {Wang}, \bibfnamefont
  {W.~Schnelle}}, \bibinfo {author} {\bibfnamefont {S.}~\bibnamefont {Wirth}},
  \bibinfo {author} {\bibfnamefont {Y.}~\bibnamefont {Chen}}, \bibinfo {author}
  {\bibfnamefont {S.}~\bibnamefont {Goennenwein}}, \ and\ \bibinfo {author}
  {\bibfnamefont {C.}~\bibnamefont {Felser}},\ }\href@noop {} {\bibfield
  {journal} {\bibinfo  {journal} {Nat. Phys.}\ }\textbf {\bibinfo {volume}
  {14}} (\bibinfo {year} {2018})}\BibitemShut {NoStop}%
\bibitem [{\citenamefont {Xu}\ \emph {et~al.}(2018)\citenamefont {Xu},
  \citenamefont {Liu}, \citenamefont {Shi}, \citenamefont {Muechler},
  \citenamefont {Gayles}, \citenamefont {Felser},\ and\ \citenamefont
  {Sun}}]{Xu2018}%
  \BibitemOpen
  \bibfield  {author} {\bibinfo {author} {\bibfnamefont {Q.}~\bibnamefont
  {Xu}}, \bibinfo {author} {\bibfnamefont {E.}~\bibnamefont {Liu}}, \bibinfo
  {author} {\bibfnamefont {W.}~\bibnamefont {Shi}}, \bibinfo {author}
  {\bibfnamefont {L.}~\bibnamefont {Muechler}}, \bibinfo {author}
  {\bibfnamefont {J.}~\bibnamefont {Gayles}}, \bibinfo {author} {\bibfnamefont
  {C.}~\bibnamefont {Felser}}, \ and\ \bibinfo {author} {\bibfnamefont
  {Y.}~\bibnamefont {Sun}},\ }\href@noop {} {\bibfield  {journal} {\bibinfo
  {journal} {Phys. Rev. B}\ }\textbf {\bibinfo {volume} {97}},\ \bibinfo
  {pages} {235416} (\bibinfo {year} {2018})}\BibitemShut {NoStop}%
\bibitem [{\citenamefont {Wang}\ \emph {et~al.}(2018)\citenamefont {Wang},
  \citenamefont {Xu}, \citenamefont {Lou}, \citenamefont {Liu}, \citenamefont
  {Li}, \citenamefont {Huang}, \citenamefont {Shen}, \citenamefont {Weng},
  \citenamefont {Wang},\ and\ \citenamefont {Lei}}]{Wang2018}%
  \BibitemOpen
  \bibfield  {author} {\bibinfo {author} {\bibfnamefont {Q.}~\bibnamefont
  {Wang}}, \bibinfo {author} {\bibfnamefont {Y.}~\bibnamefont {Xu}}, \bibinfo
  {author} {\bibfnamefont {R.}~\bibnamefont {Lou}}, \bibinfo {author}
  {\bibfnamefont {Z.}~\bibnamefont {Liu}}, \bibinfo {author} {\bibfnamefont
  {M.}~\bibnamefont {Li}}, \bibinfo {author} {\bibfnamefont {Y.}~\bibnamefont
  {Huang}}, \bibinfo {author} {\bibfnamefont {D.}~\bibnamefont {Shen}},
  \bibinfo {author} {\bibfnamefont {H.}~\bibnamefont {Weng}}, \bibinfo {author}
  {\bibfnamefont {S.}~\bibnamefont {Wang}}, \ and\ \bibinfo {author}
  {\bibfnamefont {H.}~\bibnamefont {Lei}},\ }\href@noop {} {\bibfield
  {journal} {\bibinfo  {journal} {Nat.Commun.}\ }\textbf {\bibinfo {volume}
  {9}} (\bibinfo {year} {2018})}\BibitemShut {NoStop}%
\bibitem [{\citenamefont {Liu}\ \emph {et~al.}(2019)\citenamefont {Liu},
  \citenamefont {Liang}, \citenamefont {Liu}, \citenamefont {Xu}, \citenamefont
  {Li}, \citenamefont {Chen}, \citenamefont {Pei}, \citenamefont {Shi},
  \citenamefont {Mo}, \citenamefont {Dudin}, \citenamefont {Kim}, \citenamefont
  {Cacho}, \citenamefont {Li}, \citenamefont {Sun}, \citenamefont {Yang},
  \citenamefont {Liu}, \citenamefont {Parkin}, \citenamefont {Felser},\ and\
  \citenamefont {Chen}}]{Liu2019}%
  \BibitemOpen
  \bibfield  {author} {\bibinfo {author} {\bibfnamefont {D.}~\bibnamefont
  {Liu}}, \bibinfo {author} {\bibfnamefont {A.}~\bibnamefont {Liang}}, \bibinfo
  {author} {\bibfnamefont {E.}~\bibnamefont {Liu}}, \bibinfo {author}
  {\bibfnamefont {Q.}~\bibnamefont {Xu}}, \bibinfo {author} {\bibfnamefont
  {Y.}~\bibnamefont {Li}}, \bibinfo {author} {\bibfnamefont {C.}~\bibnamefont
  {Chen}}, \bibinfo {author} {\bibfnamefont {D.}~\bibnamefont {Pei}}, \bibinfo
  {author} {\bibfnamefont {W.}~\bibnamefont {Shi}}, \bibinfo {author}
  {\bibfnamefont {S.}~\bibnamefont {Mo}}, \bibinfo {author} {\bibfnamefont
  {P.}~\bibnamefont {Dudin}}, \bibinfo {author} {\bibfnamefont
  {T.}~\bibnamefont {Kim}}, \bibinfo {author} {\bibfnamefont {C.}~\bibnamefont
  {Cacho}}, \bibinfo {author} {\bibfnamefont {G.}~\bibnamefont {Li}}, \bibinfo
  {author} {\bibfnamefont {Y.}~\bibnamefont {Sun}}, \bibinfo {author}
  {\bibfnamefont {L.}~\bibnamefont {Yang}}, \bibinfo {author} {\bibfnamefont
  {Z.~K.}\ \bibnamefont {Liu}}, \bibinfo {author} {\bibfnamefont
  {S.}~\bibnamefont {Parkin}}, \bibinfo {author} {\bibfnamefont
  {C.}~\bibnamefont {Felser}}, \ and\ \bibinfo {author} {\bibfnamefont
  {Y.}~\bibnamefont {Chen}},\ }\href@noop {} {\bibfield  {journal} {\bibinfo
  {journal} {Science}\ }\textbf {\bibinfo {volume} {365}},\ \bibinfo {pages}
  {1282} (\bibinfo {year} {2019})}\BibitemShut {NoStop}%
\bibitem [{\citenamefont {Tanaka}\ \emph {et~al.}(2020)\citenamefont {Tanaka},
  \citenamefont {Fujishiro}, \citenamefont {Mogi}, \citenamefont {Kaneko},
  \citenamefont {Yokosawa}, \citenamefont {Kanazawa}, \citenamefont {Minami},
  \citenamefont {Koretsune}, \citenamefont {Arita}, \citenamefont {Tarucha},
  \citenamefont {Yamamoto},\ and\ \citenamefont {Tokura}}]{tanaka2020}%
  \BibitemOpen
  \bibfield  {author} {\bibinfo {author} {\bibfnamefont {M.}~\bibnamefont
  {Tanaka}}, \bibinfo {author} {\bibfnamefont {Y.}~\bibnamefont {Fujishiro}},
  \bibinfo {author} {\bibfnamefont {M.}~\bibnamefont {Mogi}}, \bibinfo {author}
  {\bibfnamefont {Y.}~\bibnamefont {Kaneko}}, \bibinfo {author} {\bibfnamefont
  {T.}~\bibnamefont {Yokosawa}}, \bibinfo {author} {\bibfnamefont
  {N.}~\bibnamefont {Kanazawa}}, \bibinfo {author} {\bibfnamefont
  {S.}~\bibnamefont {Minami}}, \bibinfo {author} {\bibfnamefont
  {T.}~\bibnamefont {Koretsune}}, \bibinfo {author} {\bibfnamefont
  {R.}~\bibnamefont {Arita}}, \bibinfo {author} {\bibfnamefont
  {S.}~\bibnamefont {Tarucha}}, \bibinfo {author} {\bibfnamefont
  {M.}~\bibnamefont {Yamamoto}}, \ and\ \bibinfo {author} {\bibfnamefont
  {Y.}~\bibnamefont {Tokura}},\ }\href@noop {} {\bibfield  {journal} {\bibinfo
  {journal} {Nano Lett.}\ }\textbf {\bibinfo {volume} {20}},\ \bibinfo {pages}
  {7476} (\bibinfo {year} {2020})}\BibitemShut {NoStop}%
\bibitem [{\citenamefont {Wan}\ \emph {et~al.}(2011)\citenamefont {Wan},
  \citenamefont {Turner}, \citenamefont {Vishwanath},\ and\ \citenamefont
  {Savrasov}}]{Wan2011}%
  \BibitemOpen
  \bibfield  {author} {\bibinfo {author} {\bibfnamefont {X.}~\bibnamefont
  {Wan}}, \bibinfo {author} {\bibfnamefont {A.~M.}\ \bibnamefont {Turner}},
  \bibinfo {author} {\bibfnamefont {A.}~\bibnamefont {Vishwanath}}, \ and\
  \bibinfo {author} {\bibfnamefont {S.~Y.}\ \bibnamefont {Savrasov}},\
  }\href@noop {} {\bibfield  {journal} {\bibinfo  {journal} {Phys. Rev. B}\
  }\textbf {\bibinfo {volume} {83}},\ \bibinfo {pages} {205101} (\bibinfo
  {year} {2011})}\BibitemShut {NoStop}%
\bibitem [{\citenamefont {Burkov}\ and\ \citenamefont
  {Balents}(2011)}]{Burkov2011}%
  \BibitemOpen
  \bibfield  {author} {\bibinfo {author} {\bibfnamefont {A.~A.}\ \bibnamefont
  {Burkov}}\ and\ \bibinfo {author} {\bibfnamefont {L.}~\bibnamefont
  {Balents}},\ }\href@noop {} {\bibfield  {journal} {\bibinfo  {journal} {Phys.
  Rev. Lett.}\ }\textbf {\bibinfo {volume} {107}},\ \bibinfo {pages} {127205}
  (\bibinfo {year} {2011})}\BibitemShut {NoStop}%
\bibitem [{\citenamefont {Armitage}\ \emph {et~al.}(2018)\citenamefont
  {Armitage}, \citenamefont {Mele},\ and\ \citenamefont
  {Vishwanath}}]{Armitage2018}%
  \BibitemOpen
  \bibfield  {author} {\bibinfo {author} {\bibfnamefont {N.~P.}\ \bibnamefont
  {Armitage}}, \bibinfo {author} {\bibfnamefont {E.~J.}\ \bibnamefont {Mele}},
  \ and\ \bibinfo {author} {\bibfnamefont {A.}~\bibnamefont {Vishwanath}},\
  }\href@noop {} {\bibfield  {journal} {\bibinfo  {journal} {Rev. Mod. Phys.}\
  }\textbf {\bibinfo {volume} {90}},\ \bibinfo {pages} {015001} (\bibinfo
  {year} {2018})}\BibitemShut {NoStop}%
\bibitem [{\citenamefont {Barros}\ \emph {et~al.}(2014)\citenamefont {Barros},
  \citenamefont {Venderbos}, \citenamefont {Chern},\ and\ \citenamefont
  {Batista}}]{Barros2014}%
  \BibitemOpen
  \bibfield  {author} {\bibinfo {author} {\bibfnamefont {K.}~\bibnamefont
  {Barros}}, \bibinfo {author} {\bibfnamefont {J.~W.~F.}\ \bibnamefont
  {Venderbos}}, \bibinfo {author} {\bibfnamefont {G.-W.}\ \bibnamefont
  {Chern}}, \ and\ \bibinfo {author} {\bibfnamefont {C.~D.}\ \bibnamefont
  {Batista}},\ }\href@noop {} {\bibfield  {journal} {\bibinfo  {journal} {Phys.
  Rev. B}\ }\textbf {\bibinfo {volume} {90}},\ \bibinfo {pages} {245119}
  (\bibinfo {year} {2014})}\BibitemShut {NoStop}%
\bibitem [{\citenamefont {Ikeda}\ \emph {et~al.}(2021)\citenamefont {Ikeda},
  \citenamefont {Fujiwara}, \citenamefont {Shiogai}, \citenamefont {Seki},
  \citenamefont {Nomura}, \citenamefont {Takanashi},\ and\ \citenamefont
  {Tsukazaki}}]{Ikeda2021}%
  \BibitemOpen
  \bibfield  {author} {\bibinfo {author} {\bibfnamefont {J.}~\bibnamefont
  {Ikeda}}, \bibinfo {author} {\bibfnamefont {K.}~\bibnamefont {Fujiwara}},
  \bibinfo {author} {\bibfnamefont {J.}~\bibnamefont {Shiogai}}, \bibinfo
  {author} {\bibfnamefont {T.}~\bibnamefont {Seki}}, \bibinfo {author}
  {\bibfnamefont {K.}~\bibnamefont {Nomura}}, \bibinfo {author} {\bibfnamefont
  {K.}~\bibnamefont {Takanashi}}, \ and\ \bibinfo {author} {\bibfnamefont
  {A.}~\bibnamefont {Tsukazaki}},\ }\href@noop {} {\bibfield  {journal}
  {\bibinfo  {journal} {Commun. Mater.}\ }\textbf {\bibinfo {volume} {2}},\
  \bibinfo {pages} {1} (\bibinfo {year} {2021})}\BibitemShut {NoStop}%
\bibitem [{\citenamefont {Shiogai}\ \emph {et~al.}(2021)\citenamefont
  {Shiogai}, \citenamefont {Ikeda}, \citenamefont {Fujiwara}, \citenamefont
  {Seki}, \citenamefont {Takanashi},\ and\ \citenamefont
  {Tsukazaki}}]{Shiogai2021}%
  \BibitemOpen
  \bibfield  {author} {\bibinfo {author} {\bibfnamefont {J.}~\bibnamefont
  {Shiogai}}, \bibinfo {author} {\bibfnamefont {J.}~\bibnamefont {Ikeda}},
  \bibinfo {author} {\bibfnamefont {K.}~\bibnamefont {Fujiwara}}, \bibinfo
  {author} {\bibfnamefont {T.}~\bibnamefont {Seki}}, \bibinfo {author}
  {\bibfnamefont {K.}~\bibnamefont {Takanashi}}, \ and\ \bibinfo {author}
  {\bibfnamefont {A.}~\bibnamefont {Tsukazaki}},\ }\href@noop {} {\bibfield
  {journal} {\bibinfo  {journal} {Phys. Rev. Mater.}\ }\textbf {\bibinfo
  {volume} {5}},\ \bibinfo {pages} {024403} (\bibinfo {year}
  {2021})}\BibitemShut {NoStop}%
\bibitem [{\citenamefont {Guguchia}\ \emph {et~al.}(2020)\citenamefont
  {Guguchia}, \citenamefont {Verezhak}, \citenamefont {Gawryluk}, \citenamefont
  {Tsirkin}, \citenamefont {Yin}, \citenamefont {Belopolski}, \citenamefont
  {Zhou}, \citenamefont {Simutis}, \citenamefont {Zhang}, \citenamefont
  {Cochran}, \citenamefont {Chang}, \citenamefont {Pomjakushina}, \citenamefont
  {Keller}, \citenamefont {Wang}, \citenamefont {Lei}, \citenamefont {Amato},
  \citenamefont {Jia}, \citenamefont {Luetkens},\ and\ \citenamefont
  {Hasan}}]{Guguchia2020}%
  \BibitemOpen
  \bibfield  {author} {\bibinfo {author} {\bibfnamefont {Z.}~\bibnamefont
  {Guguchia}}, \bibinfo {author} {\bibfnamefont {J.}~\bibnamefont {Verezhak}},
  \bibinfo {author} {\bibfnamefont {D.}~\bibnamefont {Gawryluk}}, \bibinfo
  {author} {\bibfnamefont {S.}~\bibnamefont {Tsirkin}}, \bibinfo {author}
  {\bibfnamefont {J.-X.}\ \bibnamefont {Yin}}, \bibinfo {author} {\bibfnamefont
  {I.}~\bibnamefont {Belopolski}}, \bibinfo {author} {\bibfnamefont
  {H.}~\bibnamefont {Zhou}}, \bibinfo {author} {\bibfnamefont {G.}~\bibnamefont
  {Simutis}}, \bibinfo {author} {\bibfnamefont {S.-S.}\ \bibnamefont {Zhang}},
  \bibinfo {author} {\bibfnamefont {T.}~\bibnamefont {Cochran}}, \bibinfo
  {author} {\bibfnamefont {G.}~\bibnamefont {Chang}}, \bibinfo {author}
  {\bibfnamefont {E.}~\bibnamefont {Pomjakushina}}, \bibinfo {author}
  {\bibfnamefont {Z.}~\bibnamefont {Keller}, \bibfnamefont {L.~Skrzeczkowska}},
  \bibinfo {author} {\bibfnamefont {Q.}~\bibnamefont {Wang}}, \bibinfo {author}
  {\bibfnamefont {R.}~\bibnamefont {Lei}, \bibfnamefont {H.C.~Khasanov}},
  \bibinfo {author} {\bibfnamefont {A.}~\bibnamefont {Amato}}, \bibinfo
  {author} {\bibfnamefont {T.}~\bibnamefont {Jia}, \bibfnamefont {A.~Neupert}},
  \bibinfo {author} {\bibfnamefont {H.}~\bibnamefont {Luetkens}}, \ and\
  \bibinfo {author} {\bibfnamefont {M.}~\bibnamefont {Hasan}},\ }\href@noop {}
  {\bibfield  {journal} {\bibinfo  {journal} {Nat. Commun.}\ }\textbf {\bibinfo
  {volume} {11}} (\bibinfo {year} {2020})}\BibitemShut {NoStop}%
\bibitem [{\citenamefont {Guguchia}\ \emph {et~al.}(2021)\citenamefont
  {Guguchia}, \citenamefont {Zhou}, \citenamefont {Wang}, \citenamefont {Yin},
  \citenamefont {Mielke}, \citenamefont {Tsirkin}, \citenamefont {Belopolski},
  \citenamefont {Zhang}, \citenamefont {Cochran}, \citenamefont {Neupert},
  \citenamefont {Khasanov}, \citenamefont {Amato}, \citenamefont {Jia},\ and\
  \citenamefont {Luetkens}}]{Guguchia2021}%
  \BibitemOpen
  \bibfield  {author} {\bibinfo {author} {\bibfnamefont {Z.}~\bibnamefont
  {Guguchia}}, \bibinfo {author} {\bibfnamefont {H.}~\bibnamefont {Zhou}},
  \bibinfo {author} {\bibfnamefont {C.}~\bibnamefont {Wang}}, \bibinfo {author}
  {\bibfnamefont {J.-X.}\ \bibnamefont {Yin}}, \bibinfo {author} {\bibfnamefont
  {C.}~\bibnamefont {Mielke}}, \bibinfo {author} {\bibfnamefont
  {S.}~\bibnamefont {Tsirkin}}, \bibinfo {author} {\bibfnamefont
  {I.}~\bibnamefont {Belopolski}}, \bibinfo {author} {\bibfnamefont {S.-S.}\
  \bibnamefont {Zhang}}, \bibinfo {author} {\bibfnamefont {T.}~\bibnamefont
  {Cochran}}, \bibinfo {author} {\bibfnamefont {T.}~\bibnamefont {Neupert}},
  \bibinfo {author} {\bibfnamefont {R.}~\bibnamefont {Khasanov}}, \bibinfo
  {author} {\bibfnamefont {A.}~\bibnamefont {Amato}}, \bibinfo {author}
  {\bibfnamefont {M.}~\bibnamefont {Jia}, \bibfnamefont {S.~Hasan}}, \ and\
  \bibinfo {author} {\bibfnamefont {H.}~\bibnamefont {Luetkens}},\ }\href@noop
  {} {\bibfield  {journal} {\bibinfo  {journal} {npj Quantum Mater.}\ }\textbf
  {\bibinfo {volume} {6}},\ \bibinfo {pages} {1} (\bibinfo {year}
  {2021})}\BibitemShut {NoStop}%
\bibitem [{\citenamefont {Zhang}\ \emph {et~al.}(2021)\citenamefont {Zhang},
  \citenamefont {Okamoto}, \citenamefont {Samolyuk}, \citenamefont {Stone},
  \citenamefont {Kolesnikov}, \citenamefont {Xue}, \citenamefont {Yan},
  \citenamefont {McGuire}, \citenamefont {Mandrus},\ and\ \citenamefont
  {Tennant}}]{Zhang2021}%
  \BibitemOpen
  \bibfield  {author} {\bibinfo {author} {\bibfnamefont {Q.}~\bibnamefont
  {Zhang}}, \bibinfo {author} {\bibfnamefont {S.}~\bibnamefont {Okamoto}},
  \bibinfo {author} {\bibfnamefont {G.~D.}\ \bibnamefont {Samolyuk}}, \bibinfo
  {author} {\bibfnamefont {M.~B.}\ \bibnamefont {Stone}}, \bibinfo {author}
  {\bibfnamefont {A.~I.}\ \bibnamefont {Kolesnikov}}, \bibinfo {author}
  {\bibfnamefont {R.}~\bibnamefont {Xue}}, \bibinfo {author} {\bibfnamefont
  {J.}~\bibnamefont {Yan}}, \bibinfo {author} {\bibfnamefont {M.~A.}\
  \bibnamefont {McGuire}}, \bibinfo {author} {\bibfnamefont {D.}~\bibnamefont
  {Mandrus}}, \ and\ \bibinfo {author} {\bibfnamefont {D.~A.}\ \bibnamefont
  {Tennant}},\ }\href@noop {} {\bibfield  {journal} {\bibinfo  {journal} {Phys.
  Rev. Lett.}\ }\textbf {\bibinfo {volume} {127}},\ \bibinfo {pages} {117201}
  (\bibinfo {year} {2021})}\BibitemShut {NoStop}%
\bibitem [{\citenamefont {Yoshida}(1996)}]{Yoshida}%
  \BibitemOpen
  \bibfield  {author} {\bibinfo {author} {\bibfnamefont {K.}~\bibnamefont
  {Yoshida}},\ }in\ \href@noop {} {\emph {\bibinfo {booktitle} {{\it Theory of
  Magnetism}}}}\ (\bibinfo  {publisher} {Springer Series in Solid-State
  Sciences},\ \bibinfo {year} {1996})\BibitemShut {NoStop}%
\bibitem [{\citenamefont {Ozawa}\ and\ \citenamefont
  {Nomura}(2019)}]{Ozawa2019}%
  \BibitemOpen
  \bibfield  {author} {\bibinfo {author} {\bibfnamefont {A.}~\bibnamefont
  {Ozawa}}\ and\ \bibinfo {author} {\bibfnamefont {K.}~\bibnamefont {Nomura}},\
  }\href@noop {} {\bibfield  {journal} {\bibinfo  {journal} {J. Phys. Soc.
  Jpn.}\ }\textbf {\bibinfo {volume} {88}},\ \bibinfo {pages} {123703}
  (\bibinfo {year} {2019})}\BibitemShut {NoStop}%
\bibitem [{\citenamefont {Kane}\ and\ \citenamefont {Mele}(2005)}]{Kane2005}%
  \BibitemOpen
  \bibfield  {author} {\bibinfo {author} {\bibfnamefont {C.~L.}\ \bibnamefont
  {Kane}}\ and\ \bibinfo {author} {\bibfnamefont {E.~J.}\ \bibnamefont
  {Mele}},\ }\href@noop {} {\bibfield  {journal} {\bibinfo  {journal} {Phys.
  Rev. Lett.}\ }\textbf {\bibinfo {volume} {95}},\ \bibinfo {pages} {226801}
  (\bibinfo {year} {2005})}\BibitemShut {NoStop}%
\bibitem [{\citenamefont {Guo}\ and\ \citenamefont {Franz}(2009)}]{Guo2009}%
  \BibitemOpen
  \bibfield  {author} {\bibinfo {author} {\bibfnamefont {H.-M.}\ \bibnamefont
  {Guo}}\ and\ \bibinfo {author} {\bibfnamefont {M.}~\bibnamefont {Franz}},\
  }\href@noop {} {\bibfield  {journal} {\bibinfo  {journal} {Phys. Rev. B}\
  }\textbf {\bibinfo {volume} {80}},\ \bibinfo {pages} {113102} (\bibinfo
  {year} {2009})}\BibitemShut {NoStop}%
\bibitem [{\citenamefont {Luo}\ \emph {et~al.}(2021)\citenamefont {Luo},
  \citenamefont {Nakamura}, \citenamefont {Park},\ and\ \citenamefont
  {Yoon}}]{Luo2021}%
  \BibitemOpen
  \bibfield  {author} {\bibinfo {author} {\bibfnamefont {W.}~\bibnamefont
  {Luo}}, \bibinfo {author} {\bibfnamefont {Y.}~\bibnamefont {Nakamura}},
  \bibinfo {author} {\bibfnamefont {J.}~\bibnamefont {Park}}, \ and\ \bibinfo
  {author} {\bibfnamefont {M.}~\bibnamefont {Yoon}},\ }\href@noop {} {\bibfield
   {journal} {\bibinfo  {journal} {npj Comput. Mater.}\ }\textbf {\bibinfo
  {volume} {7}} (\bibinfo {year} {2021})}\BibitemShut {NoStop}%
\bibitem [{\citenamefont {Yanagi}\ \emph {et~al.}(2021)\citenamefont {Yanagi},
  \citenamefont {Ikeda}, \citenamefont {Fujiwara}, \citenamefont {Nomura},
  \citenamefont {Tsukazaki},\ and\ \citenamefont {Suzuki}}]{Yanagi2021}%
  \BibitemOpen
  \bibfield  {author} {\bibinfo {author} {\bibfnamefont {Y.}~\bibnamefont
  {Yanagi}}, \bibinfo {author} {\bibfnamefont {J.}~\bibnamefont {Ikeda}},
  \bibinfo {author} {\bibfnamefont {K.}~\bibnamefont {Fujiwara}}, \bibinfo
  {author} {\bibfnamefont {K.}~\bibnamefont {Nomura}}, \bibinfo {author}
  {\bibfnamefont {A.}~\bibnamefont {Tsukazaki}}, \ and\ \bibinfo {author}
  {\bibfnamefont {M.-T.}\ \bibnamefont {Suzuki}},\ }\href@noop {} {\bibfield
  {journal} {\bibinfo  {journal} {Phys. Rev. B}\ }\textbf {\bibinfo {volume}
  {103}},\ \bibinfo {pages} {205112} (\bibinfo {year} {2021})}\BibitemShut
  {NoStop}%
\bibitem [{\citenamefont {Grohol}\ \emph {et~al.}(2005)\citenamefont {Grohol},
  \citenamefont {Matan}, \citenamefont {Cho}, \citenamefont {Lee},
  \citenamefont {Lynn}, \citenamefont {Nocera},\ and\ \citenamefont
  {Lee}}]{Grohol2005}%
  \BibitemOpen
  \bibfield  {author} {\bibinfo {author} {\bibfnamefont {D.}~\bibnamefont
  {Grohol}}, \bibinfo {author} {\bibfnamefont {K.}~\bibnamefont {Matan}},
  \bibinfo {author} {\bibfnamefont {J.-H.}\ \bibnamefont {Cho}}, \bibinfo
  {author} {\bibfnamefont {S.-H.}\ \bibnamefont {Lee}}, \bibinfo {author}
  {\bibfnamefont {J.~W.}\ \bibnamefont {Lynn}}, \bibinfo {author}
  {\bibfnamefont {D.~G.}\ \bibnamefont {Nocera}}, \ and\ \bibinfo {author}
  {\bibfnamefont {Y.~S.}\ \bibnamefont {Lee}},\ }\href@noop {} {\bibfield
  {journal} {\bibinfo  {journal} {Nat. Mater.}\ }\textbf {\bibinfo {volume}
  {4}},\ \bibinfo {pages} {323} (\bibinfo {year} {2005})}\BibitemShut {NoStop}%
\bibitem [{\citenamefont {Kassem}\ \emph {et~al.}(2017)\citenamefont {Kassem},
  \citenamefont {Tabata}, \citenamefont {Waki},\ and\ \citenamefont
  {Nakamura}}]{Kassem2017}%
  \BibitemOpen
  \bibfield  {author} {\bibinfo {author} {\bibfnamefont {M.~A.}\ \bibnamefont
  {Kassem}}, \bibinfo {author} {\bibfnamefont {Y.}~\bibnamefont {Tabata}},
  \bibinfo {author} {\bibfnamefont {T.}~\bibnamefont {Waki}}, \ and\ \bibinfo
  {author} {\bibfnamefont {H.}~\bibnamefont {Nakamura}},\ }\href@noop {}
  {\bibfield  {journal} {\bibinfo  {journal} {Phys. Rev. B}\ }\textbf {\bibinfo
  {volume} {96}},\ \bibinfo {pages} {014429} (\bibinfo {year}
  {2017})}\BibitemShut {NoStop}%
\bibitem [{\citenamefont {Weihrich}\ and\ \citenamefont
  {Anusca}(2006)}]{Weihrich2006}%
  \BibitemOpen
  \bibfield  {author} {\bibinfo {author} {\bibfnamefont {R.}~\bibnamefont
  {Weihrich}}\ and\ \bibinfo {author} {\bibfnamefont {I.}~\bibnamefont
  {Anusca}},\ }\href@noop {} {\bibfield  {journal} {\bibinfo  {journal} {Z.
  Anorg. Allg. Chem.}\ }\textbf {\bibinfo {volume} {632}},\ \bibinfo {pages}
  {1531} (\bibinfo {year} {2006})}\BibitemShut {NoStop}%
\bibitem [{\citenamefont {Kassem}\ \emph {et~al.}(2016)\citenamefont {Kassem},
  \citenamefont {Tabata}, \citenamefont {Waki},\ and\ \citenamefont
  {Nakamura}}]{Kassem2016}%
  \BibitemOpen
  \bibfield  {author} {\bibinfo {author} {\bibfnamefont {M.~A.}\ \bibnamefont
  {Kassem}}, \bibinfo {author} {\bibfnamefont {Y.}~\bibnamefont {Tabata}},
  \bibinfo {author} {\bibfnamefont {T.}~\bibnamefont {Waki}}, \ and\ \bibinfo
  {author} {\bibfnamefont {H.}~\bibnamefont {Nakamura}},\ }\href@noop {}
  {\bibfield  {journal} {\bibinfo  {journal} {J. Solid State Chem.}\ }\textbf
  {\bibinfo {volume} {233}},\ \bibinfo {pages} {8} (\bibinfo {year}
  {2016})}\BibitemShut {NoStop}%
\bibitem [{\citenamefont {Zhou}\ \emph {et~al.}(2020)\citenamefont {Zhou},
  \citenamefont {Chang}, \citenamefont {Wang}, \citenamefont {Gui},
  \citenamefont {Xu}, \citenamefont {Yin}, \citenamefont {Guguchia},
  \citenamefont {Zhang}, \citenamefont {Chang}, \citenamefont {Lin},
  \citenamefont {Xie}, \citenamefont {Hasan},\ and\ \citenamefont
  {Jia}}]{Huibin2020}%
  \BibitemOpen
  \bibfield  {author} {\bibinfo {author} {\bibfnamefont {H.}~\bibnamefont
  {Zhou}}, \bibinfo {author} {\bibfnamefont {G.}~\bibnamefont {Chang}},
  \bibinfo {author} {\bibfnamefont {G.}~\bibnamefont {Wang}}, \bibinfo {author}
  {\bibfnamefont {X.}~\bibnamefont {Gui}}, \bibinfo {author} {\bibfnamefont
  {X.}~\bibnamefont {Xu}}, \bibinfo {author} {\bibfnamefont {J.-X.}\
  \bibnamefont {Yin}}, \bibinfo {author} {\bibfnamefont {Z.}~\bibnamefont
  {Guguchia}}, \bibinfo {author} {\bibfnamefont {S.~S.}\ \bibnamefont {Zhang}},
  \bibinfo {author} {\bibfnamefont {T.-R.}\ \bibnamefont {Chang}}, \bibinfo
  {author} {\bibfnamefont {H.}~\bibnamefont {Lin}}, \bibinfo {author}
  {\bibfnamefont {W.}~\bibnamefont {Xie}}, \bibinfo {author} {\bibfnamefont
  {M.~Z.}\ \bibnamefont {Hasan}}, \ and\ \bibinfo {author} {\bibfnamefont
  {S.}~\bibnamefont {Jia}},\ }\href@noop {} {\bibfield  {journal} {\bibinfo
  {journal} {Phys. Rev. B}\ }\textbf {\bibinfo {volume} {101}},\ \bibinfo
  {pages} {125121} (\bibinfo {year} {2020})}\BibitemShut {NoStop}%
\bibitem [{\citenamefont {Shen}\ \emph {et~al.}(2020)\citenamefont {Shen},
  \citenamefont {Zeng}, \citenamefont {Zhang}, \citenamefont {Sun},
  \citenamefont {Yao}, \citenamefont {Xi}, \citenamefont {Wang}, \citenamefont
  {Wu}, \citenamefont {Shen}, \citenamefont {Liu},\ and\ \citenamefont
  {E.}}]{Shen2020}%
  \BibitemOpen
  \bibfield  {author} {\bibinfo {author} {\bibfnamefont {J.}~\bibnamefont
  {Shen}}, \bibinfo {author} {\bibfnamefont {Q.}~\bibnamefont {Zeng}}, \bibinfo
  {author} {\bibfnamefont {S.}~\bibnamefont {Zhang}}, \bibinfo {author}
  {\bibfnamefont {H.}~\bibnamefont {Sun}}, \bibinfo {author} {\bibfnamefont
  {Q.}~\bibnamefont {Yao}}, \bibinfo {author} {\bibfnamefont {X.}~\bibnamefont
  {Xi}}, \bibinfo {author} {\bibfnamefont {W.}~\bibnamefont {Wang}}, \bibinfo
  {author} {\bibfnamefont {G.}~\bibnamefont {Wu}}, \bibinfo {author}
  {\bibfnamefont {B.}~\bibnamefont {Shen}}, \bibinfo {author} {\bibfnamefont
  {Q.}~\bibnamefont {Liu}}, \ and\ \bibinfo {author} {\bibfnamefont
  {L.}~\bibnamefont {E.}},\ }\href@noop {} {\bibfield  {journal} {\bibinfo
  {journal} {Adv. Funct. Mater.}\ }\textbf {\bibinfo {volume} {30}},\ \bibinfo
  {pages} {2000830} (\bibinfo {year} {2020})}\BibitemShut {NoStop}%
\bibitem [{\citenamefont {Kassem}\ \emph {et~al.}(2015)\citenamefont {Kassem},
  \citenamefont {Tabata}, \citenamefont {Waki},\ and\ \citenamefont
  {Nakamura}}]{Kassem2015}%
  \BibitemOpen
  \bibfield  {author} {\bibinfo {author} {\bibfnamefont {M.~A.}\ \bibnamefont
  {Kassem}}, \bibinfo {author} {\bibfnamefont {Y.}~\bibnamefont {Tabata}},
  \bibinfo {author} {\bibfnamefont {T.}~\bibnamefont {Waki}}, \ and\ \bibinfo
  {author} {\bibfnamefont {H.}~\bibnamefont {Nakamura}},\ }\href@noop {}
  {\bibfield  {journal} {\bibinfo  {journal} {J. Cryst. Growth}\ }\textbf
  {\bibinfo {volume} {426}},\ \bibinfo {pages} {208} (\bibinfo {year}
  {2015})}\BibitemShut {NoStop}%
\bibitem [{\citenamefont {Kubodera}\ \emph {et~al.}(2006)\citenamefont
  {Kubodera}, \citenamefont {Okabe}, \citenamefont {Kamihara},\ and\
  \citenamefont {Matoba}}]{Kubodera2006}%
  \BibitemOpen
  \bibfield  {author} {\bibinfo {author} {\bibfnamefont {T.}~\bibnamefont
  {Kubodera}}, \bibinfo {author} {\bibfnamefont {H.}~\bibnamefont {Okabe}},
  \bibinfo {author} {\bibfnamefont {Y.}~\bibnamefont {Kamihara}}, \ and\
  \bibinfo {author} {\bibfnamefont {M.}~\bibnamefont {Matoba}},\ }\href@noop {}
  {\bibfield  {journal} {\bibinfo  {journal} {Physica B}\ }\textbf {\bibinfo
  {volume} {378}},\ \bibinfo {pages} {1142} (\bibinfo {year}
  {2006})}\BibitemShut {NoStop}%
\bibitem [{\citenamefont {Thakur}\ \emph {et~al.}(2020)\citenamefont {Thakur},
  \citenamefont {Vir}, \citenamefont {Guin}, \citenamefont {Shekhar},
  \citenamefont {Weihrich}, \citenamefont {Sun}, \citenamefont {Kumar},\ and\
  \citenamefont {Felser}}]{Thakur2020}%
  \BibitemOpen
  \bibfield  {author} {\bibinfo {author} {\bibfnamefont {G.~S.}\ \bibnamefont
  {Thakur}}, \bibinfo {author} {\bibfnamefont {P.}~\bibnamefont {Vir}},
  \bibinfo {author} {\bibfnamefont {S.}~\bibnamefont {Guin}}, \bibinfo {author}
  {\bibfnamefont {C.}~\bibnamefont {Shekhar}}, \bibinfo {author} {\bibfnamefont
  {R.}~\bibnamefont {Weihrich}}, \bibinfo {author} {\bibfnamefont
  {Y.}~\bibnamefont {Sun}}, \bibinfo {author} {\bibfnamefont {N.}~\bibnamefont
  {Kumar}}, \ and\ \bibinfo {author} {\bibfnamefont {C.}~\bibnamefont
  {Felser}},\ }\href@noop {} {\bibfield  {journal} {\bibinfo  {journal} {Chem.
  Mater.}\ }\textbf {\bibinfo {volume} {32}},\ \bibinfo {pages} {1612}
  (\bibinfo {year} {2020})}\BibitemShut {NoStop}%
\bibitem [{\citenamefont {Dai}\ \emph {et~al.}(2005)\citenamefont {Dai},
  \citenamefont {Haule},\ and\ \citenamefont {Kotliar}}]{Dai2005}%
  \BibitemOpen
  \bibfield  {author} {\bibinfo {author} {\bibfnamefont {X.}~\bibnamefont
  {Dai}}, \bibinfo {author} {\bibfnamefont {K.}~\bibnamefont {Haule}}, \ and\
  \bibinfo {author} {\bibfnamefont {G.}~\bibnamefont {Kotliar}},\ }\href@noop
  {} {\bibfield  {journal} {\bibinfo  {journal} {Phys. Rev. B}\ }\textbf
  {\bibinfo {volume} {72}},\ \bibinfo {pages} {045111} (\bibinfo {year}
  {2005})}\BibitemShut {NoStop}%
\bibitem [{\citenamefont {Xiao}\ \emph {et~al.}(2010)\citenamefont {Xiao},
  \citenamefont {Chang},\ and\ \citenamefont {Niu}}]{Xiao2010}%
  \BibitemOpen
  \bibfield  {author} {\bibinfo {author} {\bibfnamefont {D.}~\bibnamefont
  {Xiao}}, \bibinfo {author} {\bibfnamefont {M.-C.}\ \bibnamefont {Chang}}, \
  and\ \bibinfo {author} {\bibfnamefont {Q.}~\bibnamefont {Niu}},\ }\href@noop
  {} {\bibfield  {journal} {\bibinfo  {journal} {Rev. Mod. Phys.}\ }\textbf
  {\bibinfo {volume} {82}},\ \bibinfo {pages} {1959} (\bibinfo {year}
  {2010})}\BibitemShut {NoStop}%
\bibitem [{\citenamefont {Thonhauser}\ \emph {et~al.}(2005)\citenamefont
  {Thonhauser}, \citenamefont {Ceresoli}, \citenamefont {Vanderbilt},\ and\
  \citenamefont {Resta}}]{Ceresoli2006}%
  \BibitemOpen
  \bibfield  {author} {\bibinfo {author} {\bibfnamefont {T.}~\bibnamefont
  {Thonhauser}}, \bibinfo {author} {\bibfnamefont {D.}~\bibnamefont
  {Ceresoli}}, \bibinfo {author} {\bibfnamefont {D.}~\bibnamefont
  {Vanderbilt}}, \ and\ \bibinfo {author} {\bibfnamefont {R.}~\bibnamefont
  {Resta}},\ }\href@noop {} {\bibfield  {journal} {\bibinfo  {journal} {Phys.
  Rev. Lett.}\ }\textbf {\bibinfo {volume} {95}},\ \bibinfo {pages} {137205}
  (\bibinfo {year} {2005})}\BibitemShut {NoStop}%
\bibitem [{\citenamefont {Ominato}\ \emph {et~al.}(2019)\citenamefont
  {Ominato}, \citenamefont {Tatsumi},\ and\ \citenamefont
  {Nomura}}]{Ominato2019}%
  \BibitemOpen
  \bibfield  {author} {\bibinfo {author} {\bibfnamefont {Y.}~\bibnamefont
  {Ominato}}, \bibinfo {author} {\bibfnamefont {S.}~\bibnamefont {Tatsumi}}, \
  and\ \bibinfo {author} {\bibfnamefont {K.}~\bibnamefont {Nomura}},\
  }\href@noop {} {\bibfield  {journal} {\bibinfo  {journal} {Phys. Rev. B}\
  }\textbf {\bibinfo {volume} {99}},\ \bibinfo {pages} {085205} (\bibinfo
  {year} {2019})}\BibitemShut {NoStop}%
\bibitem [{\citenamefont {Nagaosa}\ \emph {et~al.}(2010)\citenamefont
  {Nagaosa}, \citenamefont {Sinova}, \citenamefont {Onoda}, \citenamefont
  {MacDonald},\ and\ \citenamefont {Ong}}]{Nagaosa2010}%
  \BibitemOpen
  \bibfield  {author} {\bibinfo {author} {\bibfnamefont {N.}~\bibnamefont
  {Nagaosa}}, \bibinfo {author} {\bibfnamefont {J.}~\bibnamefont {Sinova}},
  \bibinfo {author} {\bibfnamefont {S.}~\bibnamefont {Onoda}}, \bibinfo
  {author} {\bibfnamefont {A.~H.}\ \bibnamefont {MacDonald}}, \ and\ \bibinfo
  {author} {\bibfnamefont {N.~P.}\ \bibnamefont {Ong}},\ }\href@noop {}
  {\bibfield  {journal} {\bibinfo  {journal} {Rev. Mod. Phys.}\ }\textbf
  {\bibinfo {volume} {82}},\ \bibinfo {pages} {1539} (\bibinfo {year}
  {2010})}\BibitemShut {NoStop}%
\end{thebibliography}
